\documentclass{ws-p8-50x6-00}
\usepackage{graphicx}
\usepackage{dcolumn}
\usepackage{bm}
\newcommand{\bee}{\begin{eqnarray}}
\newcommand{\eee}{\end{eqnarray}}
\newcommand{\eq}{\end{quote}}
\newcommand{\nn}{\nonumber}
\newcommand{\Slash}[1]{\ooalign{\hfil/\hfil\crcr$#1$}}

\def\lsim{\displaystyle\mathop{<}_{\sim}}
\begin{document}
\title{$\Theta^+$ baryon production from $\gamma N$ and $NN$ scattering}
\author{Seung-Il Nam}
\address{Research Center for Nuclear Physics (RCNP), Osaka University,
  Ibaraki, Osaka 567-0047, Japan and \\ Nuclear physics \& Radiation
  technology Institute (NuRI), 
  Pusan University, Keum-Jung Gu, Busan 609-735, Korea \\
  sinam@rcnp.osaka-u.ac.jp} 
\author{Atsushi Hosaka}
\address{Research Center for Nuclear Physics (RCNP), Osaka University,
  Ibaraki, Osaka 567-0047, Japan \\ Hosaka@rcnp.osaka-u.ac.jp} 
\author{Hyun-Chul Kim}
\address{Nuclear physics \& Radiation technology Institute (NuRI),
  Pusan University, Keum-Jung Gu, Busan 609-735, Korea \\ hchkim@pusan.ac.kr} 
\maketitle
\abstracts{We investigate $\Theta^+$ production via $\gamma N$
  and $NN$ reactions in order to obtain information on the structure
  of $\Theta^+$, especially its parity.  We observe that the positive parity
  $\Theta^+$ production provides about ten times larger total cross
  sections than those of the negative parity one in both photon and nucleon
  induced reactions due to $P$--wave enhancement of the $KN\Theta$
  vertex. We also consider the model independent method in the nucleon induced
  reaction to determine the  
  parity of $\Theta^+$ and show clearly distinguishable signals for the
  two parities.}
\section{Introduction}
After the observation of the evidence of $\Theta^+$ by LEPS group at
SPring-8~\cite{Nakano:2003qx} motivated by Diakonov  
{\it et al.}~\cite{Diakonov:1997mm}, exotic pentaquark baryon state $\Theta^+$
has triggered huge amounts of the research activities in both 
experimental $^{3\sim 11}$ and theoretical $^{12\sim 33}$ hadron
physics fields. Recent experimental situation is rather controversial,
and the existence of the $\Theta^+$ still needs confirmation.
However, it is strongly expected that physics of pentaquarks will open
a new challenge for hadron physics with rich structure of
non-perturbative QCD. Although the present experimental information is
limited, it is therefore, of great importance to analyze what we can
learn from the experiments done so far and in the future. In this
paper, we report the series of our works for the $\Theta^+$ production
reaction in different approaches including $\gamma N$ and $NN$ induced
ones. Our aim is to extract information of $\Theta^+$ structure,
especially the parity of $\Theta^+$. As discussed in
Refs.~\cite{Stancu:2003if,Hosaka:2003jv}, the parity of $\Theta^+$ carries important
informations of the dynamics of low energy QCD. Contents of this paper
is as follows. In section 2, the method of calculation in an
effective Lagrangian approach is briefly
formulated for the reactions, $\gamma N\to \bar{K}\Theta^+\,(N=p,n)$
and $NN\to Y\Theta^+\,(N=p,n;\,Y=\Lambda,\Sigma)$. In all cases, we
perform calculation for $J^P(\Theta^+)=1/2^+$ and $1/2^-$. For the
$pp$ induced reaction, we consider a polarized one as suggested by
Hanhart~\cite{Hanhart:2003xp} for the unambiguous determination of the parity of
$\Theta^+$. In section 3, we present numerical results and discuss
various aspects of the above reactions. Final section is devoted to
summary of the present report.   
\section{Formalism}
\subsection{$\gamma N$ scattering}
We start with an effective Lagrangian approach for the $\gamma N$
scattering for the
tree level calculations. Concerning to the $KN\Theta$ vertex, we utilize two different
interactions, {\em i.e.}, the pseudoscalar (PS) and pseudovector (PV)
schemes.  The effective Lagrangians for the reactions are 
given as follows:
\bee
\mathcal{L}_{N\Theta K}
&=&
ig\bar{\Theta}\Gamma_{5}K N + \mbox{h.c.},  
\nn\\
\mathcal{L}_{N\Theta K}
&=&
-\frac{g^{\ast}_{A}}{2f_{\pi}}
\bar{\Theta}\gamma_{\mu}\Gamma_{5}\partial^{\mu}K N + \mbox{h.c.},
\label{ntkpv} \nn\\
\mathcal{L}_{\gamma KK}
&=&
ie \left\{ K (\partial^{\mu}\bar{K}) -
  (\partial^{\mu}K)\bar{K}\right\} A_{\mu} + {\rm h.c.},
\nn\\
\mathcal{L}_{\gamma NN}
&=&
-e\bar{N}\left(\gamma_{\mu}
+i\frac{\kappa_{N}}{2M_{N}}
\sigma_{\mu\nu}k^{\nu}\right)N \, A^{\mu} + {\rm h.c.}\, 
\nn\\
\mathcal{L}_{\gamma\Theta \Theta}
&=&
-e\bar{\Theta}\left(\gamma_{\mu}
+i\frac{\kappa_{\Theta}}{2M_{\Theta}}
\sigma_{\mu\nu}k^{\nu}\right)\Theta \, A^{\mu}\ + {\rm h.c.}, 
\label{gtt}
\eee 
where $\Theta$, $N$, and $K$ stand for the pentaquark 
$\Theta^+$, the nucleon, and the kaon fields, respectively.  
Parameters $e$,
$\kappa$, and $M$ designate the electric charge, the anomalous
magnetic moment, and the mass of baryon, respectively.  $\Gamma_{5}$
is $\gamma_{5}$ for the positive-parity $\Theta^{+}$
($\Theta_{+}^{+}$) and ${\bf 1}_{4\times 4}$ for the negative-parity
$\Theta^{+}$ ($\Theta_{-}^{+}$).  In the case of the positive-parity
$\Theta^+$, the coupling constants for the $KN\Theta^+$ vertex  can be
determined by using a decay width 
$\Gamma_{\Theta\rightarrow KN} = 15\, {\rm MeV}$ and the mass
$M_{\Theta} = 1540\, {\rm MeV}$, from which we obtain 
$g^{\ast}_{A}=0.28$ for the PV interaction as well as $g=3.8$ for the
PS.  Similarly, we find $g^{\ast}_{A}=0.16$ and $g=0.53$ for the
negative-parity one. $K^{*}$ exchange is also taken into account in this work as in
Refs.~\cite{Liu:2003rh,Oh:2003kw,Liu:2003zi,Janssen:2001wk}.  The
corresponding Lagrangians are given as follows:
\bee
\mathcal{L}_{\gamma {K} K^{*}}&=& g_{\gamma K K^{*}}
\epsilon_{\mu\nu\sigma\rho}(\partial^{\mu}A^{\nu})
(\partial^{\sigma}K^{\dagger}){K}^{*\rho} 
+ {\rm h.c.},\nn\\ 
\label{gkv}
\mathcal{L}_{K^{*}N\Theta}&=&g_{K^{*}N\Theta}\bar{\Theta}
\gamma^{\mu}\bar{\Gamma}_{5}{K}^{*\dagger}_{\mu}N
+ {\rm h.c.}. 
\label{vnt}
\eee
We neglect the tensor coupling of the $K^{*}N\Theta$ vertex for the lack
of information.  In order to determine the coupling constant
$g_{\gamma K K^{*}}$, we use the experimental data for the radiative
decay, which gives $0.388\, {\rm GeV}^{-1}$ for the neutral decay and 
$0.254 \,{\rm GeV}^{-1}$ for the charged
decay~\cite{Oh:2003kw,Liu:2003zi,particle}. $\bar{\Gamma}_{5}$
denotes ${\bf 1}_{4\times 4}$ for the 
$\Theta^{+}_{+}$ and $\gamma_{5}$ for the $\Theta^{+}_{-}$.  Although we
have no information on $g_{K^{*}N\Theta}$ experimentally, we adopt
its value as $g_{K^{*}N\Theta}/g_{KN\Theta} = \pm 0.5$, assuming the ratio
similar to $g_{K^{*}N\Lambda}/g_{KN\Lambda}$.  Note that in
Refs.~\cite{Liu:2003zi,Yu:2003eq} the ratio of the couplings was taken to be 
$0.5$. In addition to $K^*$ exchange, we also consider $K_{1}(1270)$
axial-vector meson exchange.  However, since we find that its
contribution is tiny as found in Ref.~\cite{Yu:2003eq}, we will not take
into account it in this work.   
Since the anomalous magnetic moment of $\Theta^{+}$ has not been known
neither, we need to rely on the model
calculations~\cite{Nam:2003uf,Zhao:2003gs,Kim:2003ay,Huang:2003bu,Liu:2003ab}.
Many of these calculations indicate small numbers for the $\Theta^+$ 
magnetic moment and hence negative values for the 
anomalous magnetic moment.  
As a typical value, 
we shall use for the anomalous magnetic moment 
$\kappa_{\Theta} = -0.8 \mu_N$. In the PV scheme, we need to consider an additional contribution,
{\em i.e.}, the contact term, also known as the Kroll-Rudermann (KR)
term.  The term can be written as follows.
\be
i\mathcal{M}_{\rm KR,}=
-e\frac{g^{\ast}_{A}}{2f_{\pi}}\bar{u}(p')\Gamma_{5}\Slash{\epsilon}u(p)
\label{M11}.
\ee
While Yu {\em et al.}~\cite{Yu:2003eq} introduced the formar
factors into the KR term in such a way that they satisfy the gauge
invariance, we make use of the following relation:
\bee
i\Delta\mathcal{M}^{0}=i\mathcal{M}^{0}_{\rm PV}-i\mathcal{M}^{0}_{\rm PS}&=& 
e\frac{g}{M_{N}+M_{\Theta}}
\left(\frac{\kappa_{\Theta}}{2M_{\Theta}}+\frac{\kappa_{N}}{2M_{N}}
\right) 
\b{u}(p')\Gamma_{5}\Slash{\epsilon}\Slash{k}u(p).
\eee
Here, The superscript $0$ denotes the bare amplitudes without the form
factor.  
Since $i\Delta\mathcal{M}^{0}$ is gauge-invariant due to 
its tensor structure, 
we can easily insert the form factors, keeping the gauge invariance.  
Thus,  we arrive at the gauge-invariant amplitudes in 
the PV scheme as follows:
\bee
i\mathcal{M}_{\rm PV}&=&i\mathcal{M}_{\rm PS}+i\Delta\mathcal{M}\nn\\
&=&i\mathcal{M}_{\rm PS}+e\frac{g}{M_{N}+M_{\Theta}}
\left(F_{u}\frac{\kappa_{\Theta}}{2M_{\Theta}}+F_{s}\frac{\kappa_{N}}{2M_{N}}
\right) 
\b{u}(p')\Gamma_{5}\Slash{\epsilon}\Slash{k}u(p).
\label{eq:gauge}
\eee    
Finally, the $K^{*}$-exchange amplitude is derived as ar
follows:
\bee
\mathcal{M}_{K^{*}}=i\frac{F_{t}g_{\gamma K
K^{*}}g_{K^{*} N\Theta}}{(k-k')^{2}-M^{2}_{K^{*}}}\bar{u}(p')
\epsilon_{\mu\nu\sigma\rho}k^{\mu}\epsilon^{\nu}k'^{\sigma}\gamma^{\rho}
\bar{\Gamma}_{5}u(p), 
\eee 
which is clearly gauge-invariant.     
\subsection{$N N$ scattering}
In this section, we formulate $NN$ scattering with $K$ and vector
$K^{*}$ exchanges in the $t$--channel. The initial and final state
interactions are not considered here. We will discuss briefly their effect later.  As mentioned 
before, we treat the reactions in the case of positive- and
negative-parity $\Theta^{+}$.  We distinguish the positive-parity
$\Theta^+$ from the negative-parity one by expressing them as
$\Theta^{+}_{+}$ and $\Theta^{+}_{-}$, respectively.  We start with
the following effective Lagrangians.
\bee
\mathcal{L}_{KNY}&=&-ig_{KNY}\bar{Y}\gamma_{5}K^{\dagger}N,\nn\\
\label{knyl}
\mathcal{L}_{KN\Theta_{\pm}}&=&-ig_{KN\Theta_{\pm}}\bar{\Theta}_{\pm}\Gamma_{5}KN,     
\label{kntl} \cr
\mathcal{L}_{VNY}&=&-g_{VNY}\bar{Y}\gamma_{\mu}
V^{\mu}N-\frac{g^{T}_{VNY}}{M_{Y}+M_{N}}\bar{Y}
\sigma_{\mu\nu}\partial^{\nu}V^{\mu}N,\nn\\ 
\mathcal{L}_{VN\Theta}&=&-g_{VN\Theta_{\pm}}
\bar{\Theta}_{\pm}\gamma_{\mu}\bar{\Gamma}_{5}V^{\mu}
N-\frac{g^{T}_{VN\Theta_{\pm}}}{M_{\Theta}+M_{N}}
\bar{\Theta}_{\pm}\sigma_{\mu\nu}\bar{\Gamma}_{5}\partial^{\nu}V^{\mu}N,
\label{effL}
\eee  
where $Y$, $K$, $N$, $\Theta$, and $V$ stand for the hyperon
($\Sigma$ and $\Lambda$), kaon, nucleon, $\Theta^{+}$, and vector
meson fields, respectively. When their signs are the same, the 
$K^{*}N\Theta$ (magnetic) coupling strength which is the sum of the
vector and tensor couplings amounts to be $1.5|g_{KN\Theta}|$.  The
value is similar to the one estimated in a fall apart mechanism,
$g_{K^{*}N\Theta} = \sqrt{3}g_{KN\Theta}$~\cite{Close:2004tp}. 
We employ the values of the 
$KNY$ and $K^{*}NY$ coupling constants referring to those from 
the new
Nijmegen potential (averaged values of models NSC97a and
NSC97f~\cite{stokes} as well as from the J\"ulich--Bonn 
YN potential (model $\tilde{\rm A}$)~\cite{Reuber:ip} 
\subsection{Polarized $pp$ scattering}
An unambiguous method to determine the parity
of the $\Theta^+$ was proposed using the reaction~\cite{Thomas:2003ak}
\bee
\vec p + \vec p \to \Theta^+ + \Sigma^+ \; \; \; 
{\rm near}\; {\rm threshold}.  
\label{reaction}
\eee
This reaction has been previously considered for
the production of $\Theta^+$~\cite{Polyakov:1999da}, but
it has turned out that it does more for the determination of the parity,
in contrast with number of recent attempts using other reactions
which needed particular production mechanism. In order to extract
information of parity from (\ref{reaction}), 
the only requirement is that the final state is dominated by
the s-wave component.
The s-wave dominance in the final state is then combined with the
Fermi statistics of the initial two protons and conservations of
the strong interaction, establishing the selection rule:
{\em If the parity of $\Theta^+$ is positive, the reaction
(\ref{reaction}) is allowed at the threshold region
only when the two protons have the total spin 
$S = 0$ and even values of relative momenta $l$, 
while, if it is negative the reaction is allowed only when
they have $S=1$ and odd $l$ values.}
This situation is similar to what was used in determining the parity
of the pion~\cite{panofsky}.
Experimentally, the pure $S = 0$ state may not be easy to set up.
However, an appropriate combination of spin polarized quantities
allows to extract  
information of $S = 0$ state.  In Ref.~\cite{Hanhart:2003xp}, the authors discussed the
experimental methods and observable to determine the parity of $\Theta^+$ baryon 
with the polarized proton beam and target. They discussed the spin
observable $A_{xx}$ as well as cross sections. It is computed by
\be
A_{xx}=\frac{(^{3}\sigma{_0}+^{3}\sigma_{1})}{2\sigma_{0}}-1,
\label{axxeq}
\ee
where $\sigma_{0}$ is the unpolarized total cross sections and the
polarized cross section are denoted as $^{2S+1}\sigma_{S_z}$. 
\subsection{Form-factor for the extended hadron structure}
As for the $\gamma N$ scattering, we have introduced the
form factors $F_{s,u,t}$ and $F^{n}_{c}$ 
in such a way that they satisfy the gauge
invariance~\cite{Ohta:ji,Haberzettl:1998eq,Davidson:2001qs} in the
form of  
\bee
F_{\rm \xi}=\frac{\Lambda^{4}}{\Lambda^{4}+\left({\rm \xi}-M^{2}_{\rm \xi}\right)^{2}},
\label{ff}
\eee
where $\xi$ represents relevant kinematic channels, $s$, $t$, and $u$,
generically.  The common form factor $F_{\rm c}$ is introduced 
according to the prescription suggested by
Refs.~\cite{Davidson:2001qs}:    
\bee
F^{\rm n}_{\rm c}&=&F_{\rm u}+F_{\rm t}-F_{\rm u}F_{\rm t},\nn\\F^{\rm
p}_{\rm c}&=&F_{\rm s}+F_{\rm u}-F_{\rm s}F_{\rm u}. 
\eee
The cutoff parameter for Eq.~(\ref{ff}) will be given in the next
section considering $\gamma p\to K^+\Lambda(1115)$ process. As for the $NN$ scattering, in order to
compute the cross sections for these reactions, we need the form
factors at each vertex to take into 
account the extended size of hadrons.  
For the Nijmegen potential we
introduce a monopole-type form factor~\cite{Machleidt:hj} in the form of 
\bee
F(q^{2}) = \frac{\Lambda^{2}-m^{2}}{\Lambda^{2}-t},
\label{ff1}
\eee
where $m$ and $t$ are the meson mass and a squared four momentum
transfer, respectively. The value of the cutoff parameter is 
taken to be 1.0 GeV for the parameter set of the Nijmegen 
potential.  We will also employ the Nijmegen
potential with the form factor, Eq.~(\ref{ff1}) for the polarized $pp$
calculation~\cite{Nam:2004qy}. As for that of the J\"ulich--Bonn 
potential, we make use of the following form factor taken from 
Ref.~\cite{Reuber:ip}:  
\bee
F(q^{2}) = \frac{\Lambda^{2}-m^{2}}{\Lambda^{2}+|{\bf q}|^{2}},
\label{ff2}
\eee
where $|{\bf q}|$ is the three momentum transfer.  In this case, we
take different values of the cutoff masses for each $KNY$ vertex as
follows~\cite{Reuber:ip}: $\Lambda_{KN\Theta} = 
\Lambda_{K^{*}N\Theta} = 1.0\, {\rm GeV}$, $\Lambda_{KN\Lambda}=1.2\,
{\rm GeV}$,  $\Lambda_{K^{*}N\Lambda}=2.2\, {\rm GeV}$,
$\Lambda_{KN\Sigma}=2.0\,{\rm GeV}$, and
$\Lambda_{K^{*}N\Sigma}=1.07\, {\rm GeV}$. 
\section{Numerical results}
\subsection{$\gamma N$ scattering}

\begin{figure}[tbh]
\begin{tabular}{cc}
\resizebox{5.5cm}{4cm}{\includegraphics{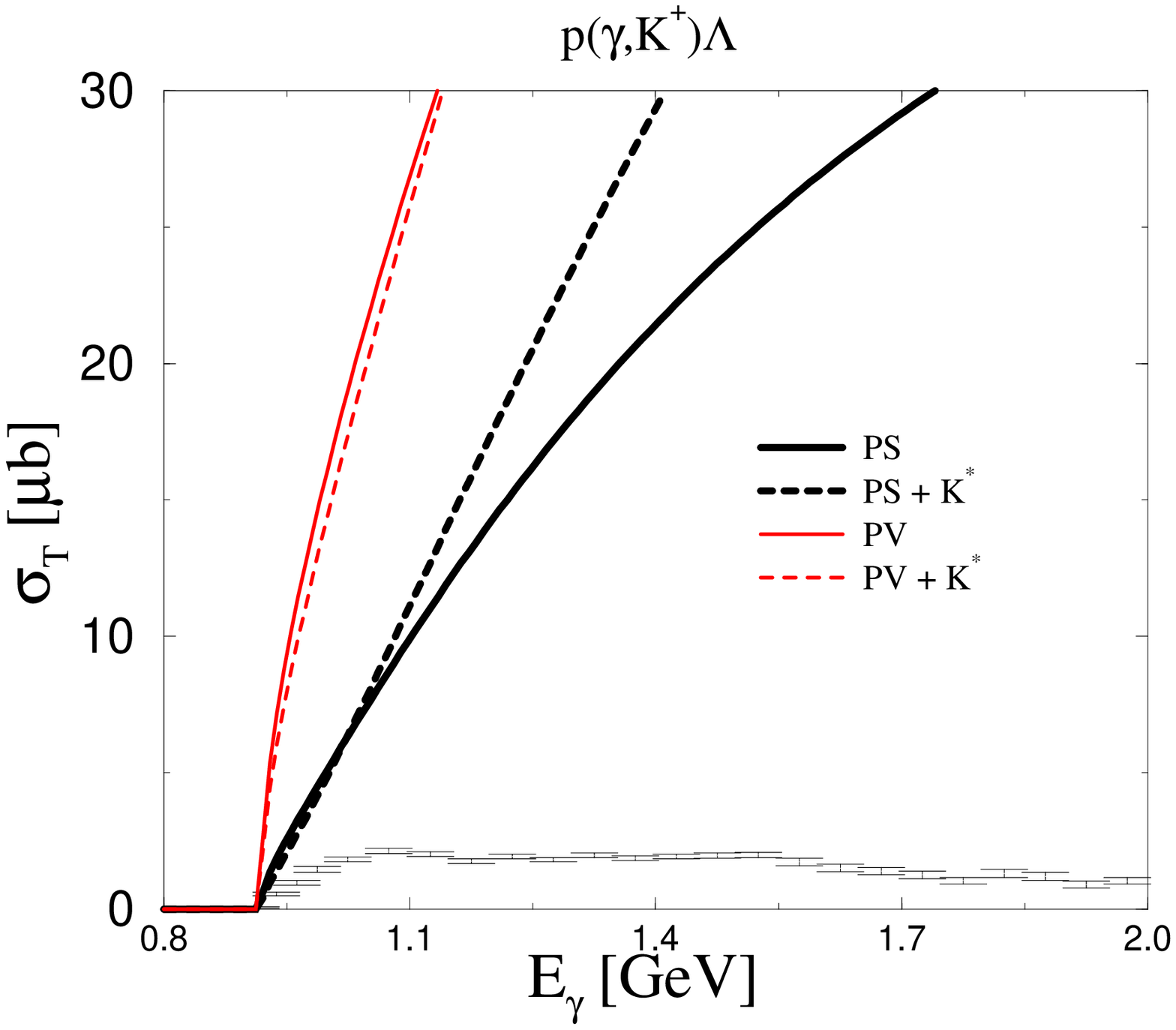}}
\resizebox{5.5cm}{4cm}{\includegraphics{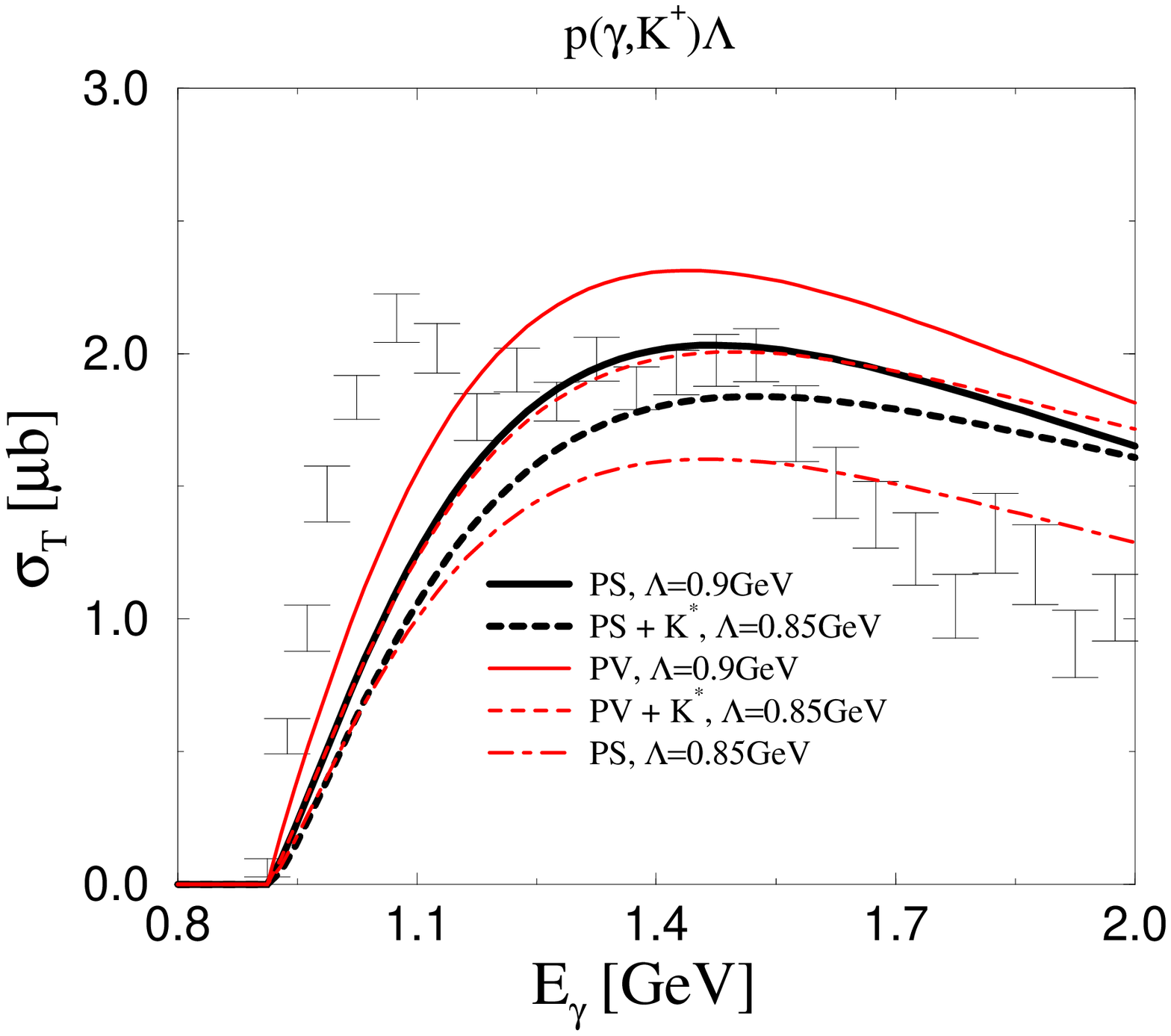}}
\end{tabular}
\caption{The total cross sections of
$\gamma p\rightarrow K^{+}\Lambda$ without (the left) and with (the right) the form
factors written in Eq.~(\ref{ff}).}
\label{gn1}
\end{figure}

Before we calculate the photoproduction of the $\Theta^+$ numerically,
we need to fix the cutoff parameters in the form factors.  In doing
so, we will try to estimate the value of the cutoff parameters by
considering the process $\gamma p\rightarrow K^{+}\Lambda$, which is
known experimentally~\cite{Tran:qw} and the comparison of the theoretical 
prediction with the corresponding data is possible.  Fig.~\ref{gn1} we
present the total cross sections of the $\gamma p\rightarrow
K^{+}\Lambda$ reaction without the form factors.  Here, we have
employed the coupling constants $g_{KN\Lambda} = -13.3$ and
$g_{K^{*}N\Lambda} = -6.65$.  While the results without form factors
are monotonically increased unphysically as shown in the left panel of
Fig.~\ref{gn1}, those with the form factors defined in
Eq.~(\ref{ff}) describe relatively well the experimental data as in
the right panel of Fig.~\ref{gn1}.  We find that $\Lambda = 0.85
\sim 0.9\,{\rm GeV}$ give reasonable results qualitatively.  Note that
the peaks at around 1.0 GeV and 1.5 GeV in the experimental data are
believed to be related to higher nucleon resonances such as
$S_{11}(1650)$, $P_{11}(1710)$, $P_{13}(1720)$ and
$D_{13}(1895)$~\cite{Janssen:2001wk}, which in our calculations are
not included.      
\begin{figure}[tbh]
\begin{tabular}{cc}
\resizebox{5.5cm}{4cm}{\includegraphics{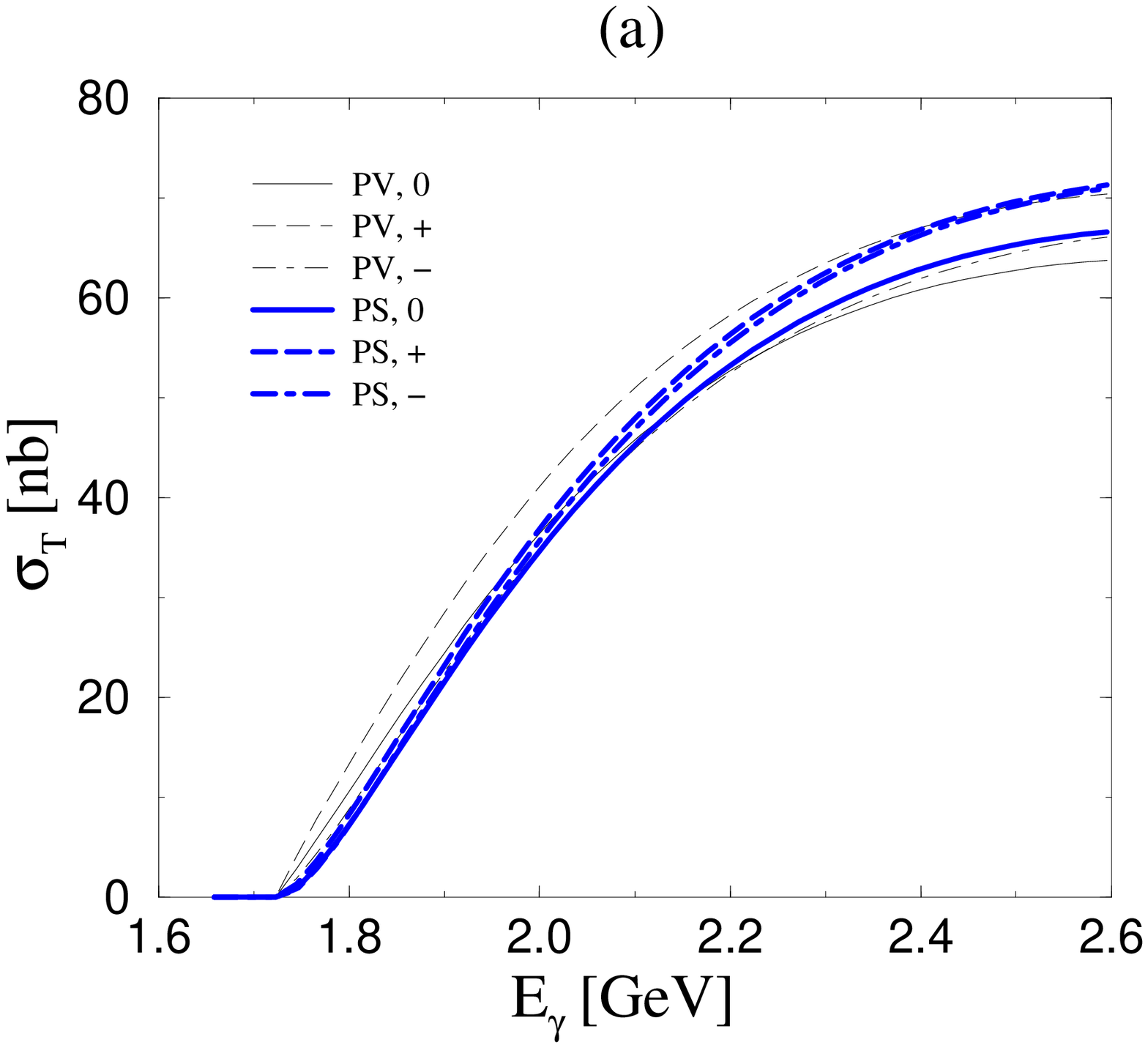}}
\resizebox{5.5cm}{4cm}{\includegraphics{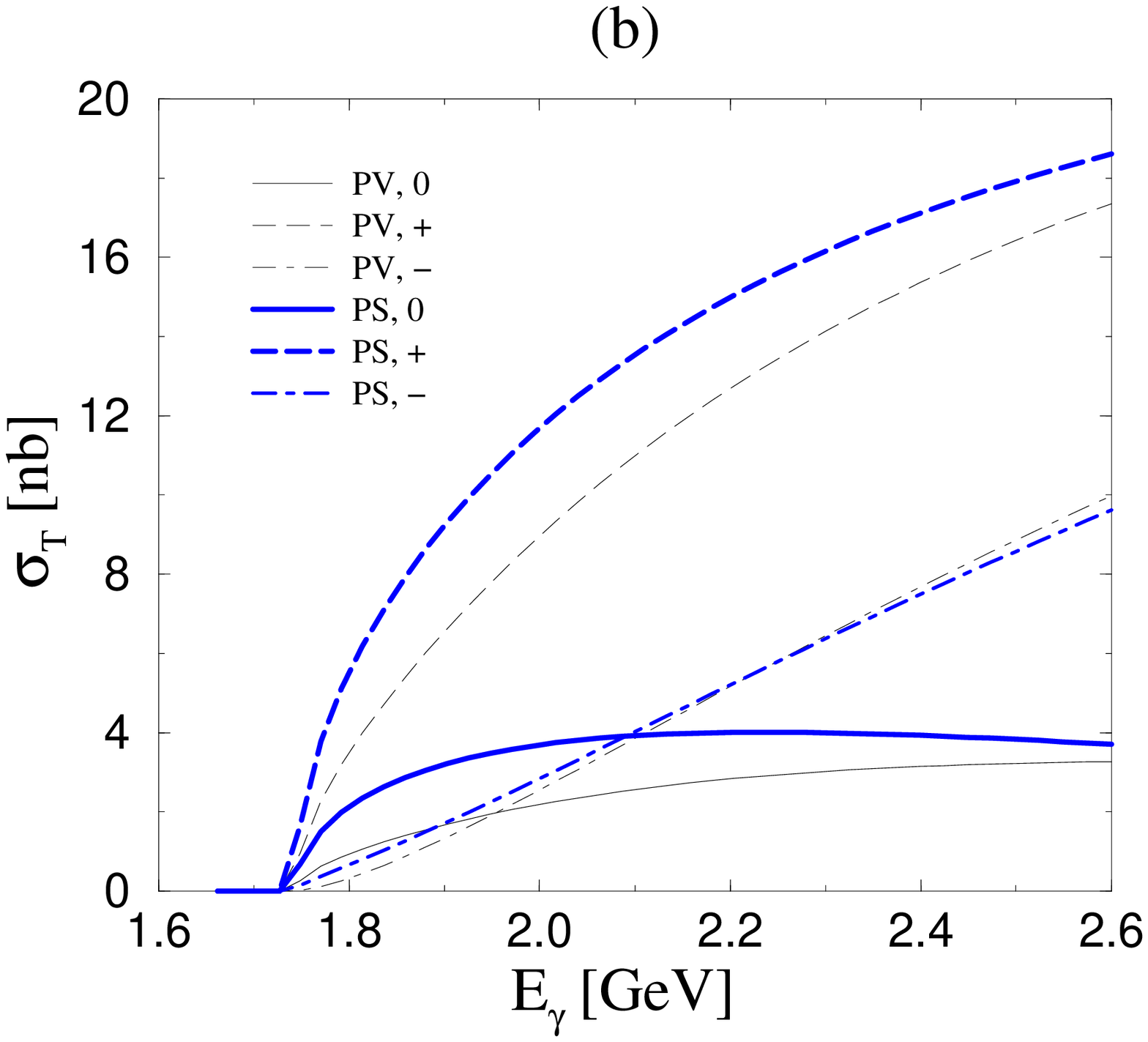}}
\end{tabular}
\caption{The total cross sections for the reactions of $\gamma
n\rightarrow K^{-}\Theta^+_{+} $
(a) and $\gamma p\rightarrow  \b{K}^{0}\Theta^+_{+}$ (b). PV and PS indicate the 
coupling schemes. 0, + and - indicate
$g_{K^{*}N\Theta}=0$, $g_{K^{*}N\Theta}=g_{KN\Theta}/2$ and
$g_{K^{*}N\Theta}=-g_{KN\Theta}/2$, respectively.}ar
\label{gn2}
\end{figure}    

Based on these results, we assume that the cutoff parameter for the
$KN\Theta$ vertex is the same as for the $KN\Lambda$ one
and use $\Lambda = 0.85$ GeV.  Fig.~\ref{gn2} shows the total
cross sections with the form factors and $g_{K^{*}N\Theta}$ being varied
between $-g_{KN\Theta}/2$ and $g_{KN\Theta}/2$.  
We see that the
differences between the PV and PS schemes turn out to be small,
as compared to the results of Ref.~\cite{Yu:2003eq}.  
The reason lies in the fact that 
Ref.~\cite{Yu:2003eq} introduced the form factor in the KR term
directly, while we employ the relation between the PV and PS
schemes as given in Eq.~(\ref{eq:gauge}).  It is very natural that in 
the low-energy limit the difference between the PV and PS schemes should
disappear.  In this sense, the present results is consistent 
with the low-energy relation for the photo-production. Coming to the
photo-production of the $\Theta^+$ in the $\gamma 
p\rightarrow \bar{K}^0 \Theta^+$ reaction, we notice that the
total cross section is smaller than the case of $\gamma n$ and 
rather sensitive to the contribution of $K^*$ exchange.  It can be
understood by the fact that the contribution of $K$ exchange is absent
and the $s$-- and the $u$--channels are suppressed by the form factors.  The
average values of the total cross sections are estimated as 
follows: $\sigma_{\gamma n\rightarrow K^-\Theta^+}\sim 44 \,{\rm nb}$
and $\sigma_{\gamma p\rightarrow \bar{K}^0\Theta^+}\sim 13 \,{\rm
  nb}$ in the range of the photon energy $1.73\,{\rm GeV} < E_{\gamma} <
2.6 \,{\rm GeV}$.  Note that these values are smaller than those of
Ref.~\cite{Nam:2003uf}, where $\Lambda = 1.0\,{\rm  GeV}$ is employed.  
\begin{figure}[tbh]
\begin{tabular}{cc}
\resizebox{5.5cm}{4cm}{\includegraphics{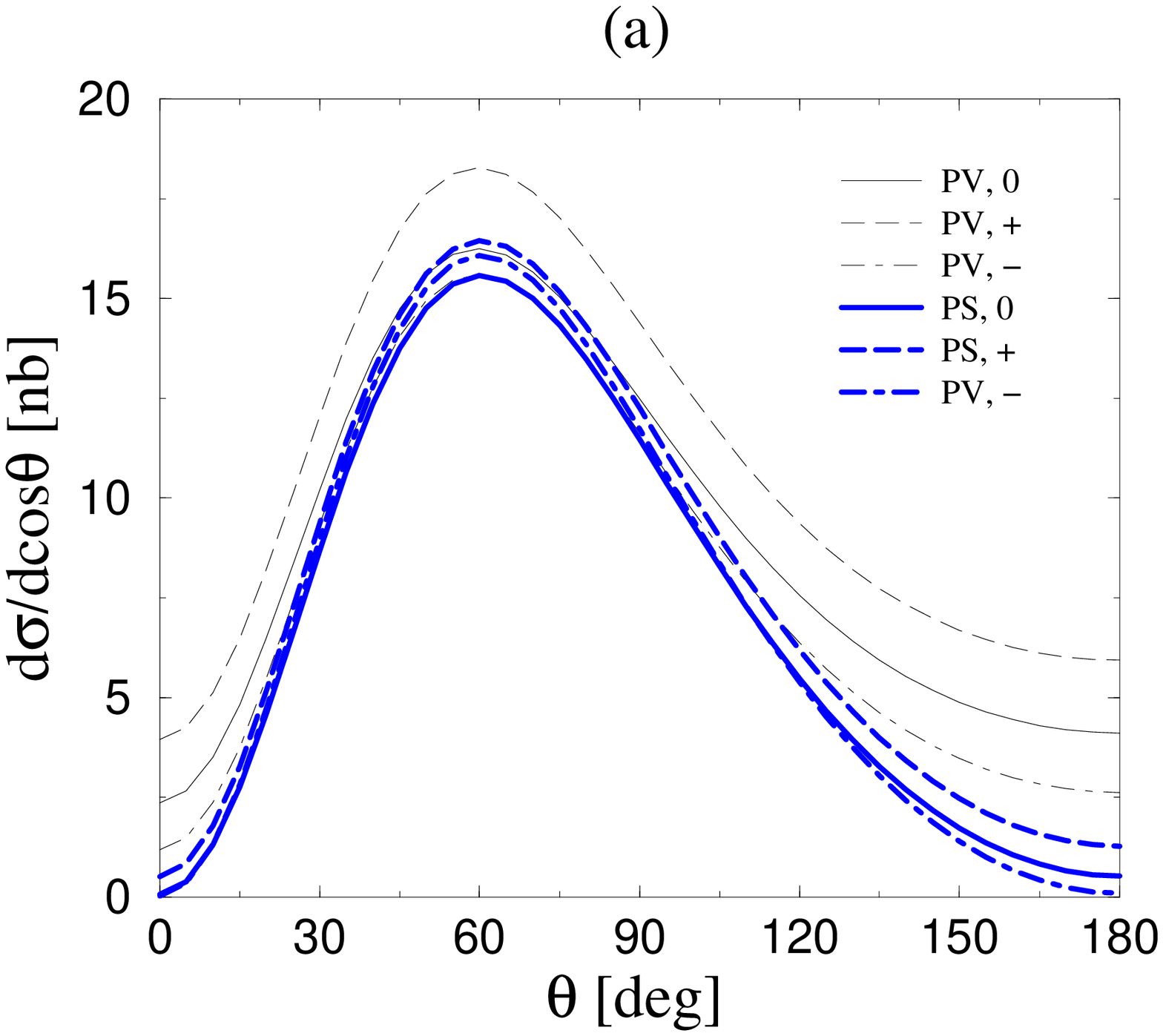}}
\resizebox{5.5cm}{4cm}{\includegraphics{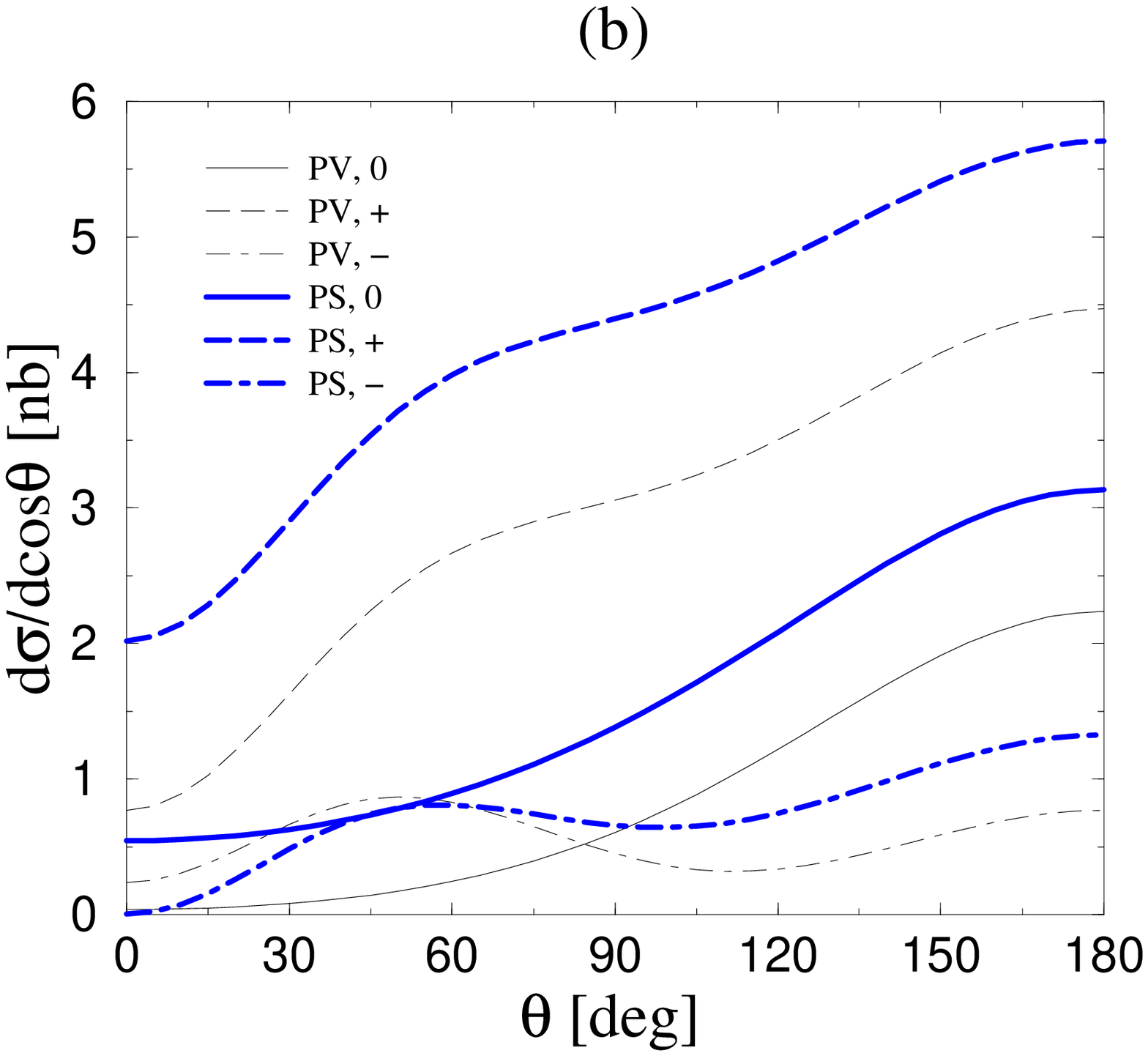}}
\end{tabular}
\caption{The differential cross sections for the reactions of $\gamma
n\rightarrow K^{-}\Theta^{+}_{+}$ (a) and $\gamma
p\rightarrow  \bar{K}^{0}\Theta^{+}_{+}$ (b) at $\sqrt{s} = 2.1\,{\rm  GeV}$.}
\label{gn3}
\end{figure} 

In Fig.~\ref{gn3}, we draw differential cross sections.  In the
case of the $\gamma n\rightarrow K^-\Theta^+$, 
the peak around $60^\circ$ is clearly seen as shown in the left panel
of Fig.~\ref{gn3}.   
This peak is caused by the $t$--channel dominance which brings about  
the combination of the factor
$|\epsilon\cdot k^{'}|^{2}\sim \sin^{2}\theta$ and the form factor.  
In the multipole basis, an $M1$ amplitude is responsible for it.  
In contrast, for the production from the proton, 
$K$ exchange is absent, and the role of $K^{*}$ exchange and its
interference with the $s$-- and the $u$--channel diagrams become more 
important.  Therefore, the differential cross section of the  
$\gamma p\rightarrow \bar{K}^0\Theta^+$ process is quite
different from that of the $\gamma n\rightarrow K^-\Theta^+$.  The
present results look rather different from those of 
Ref.~\cite{Oh:2003kw}, where the relation $g_{K^{*}N\Theta}=\pm
g_{KN\Theta}$ was employed.  It is so since the amplitude of $K^*$ 
exchange is twice as large as that in the present work, 
and has an even more important contribution to the amplitudes.  We
need more experimental information in order to settle the uncertainty
in the reaction mechanism.   

\begin{figure}[tbh]
\begin{tabular}{cc}
\resizebox{5.5cm}{4cm}{\includegraphics{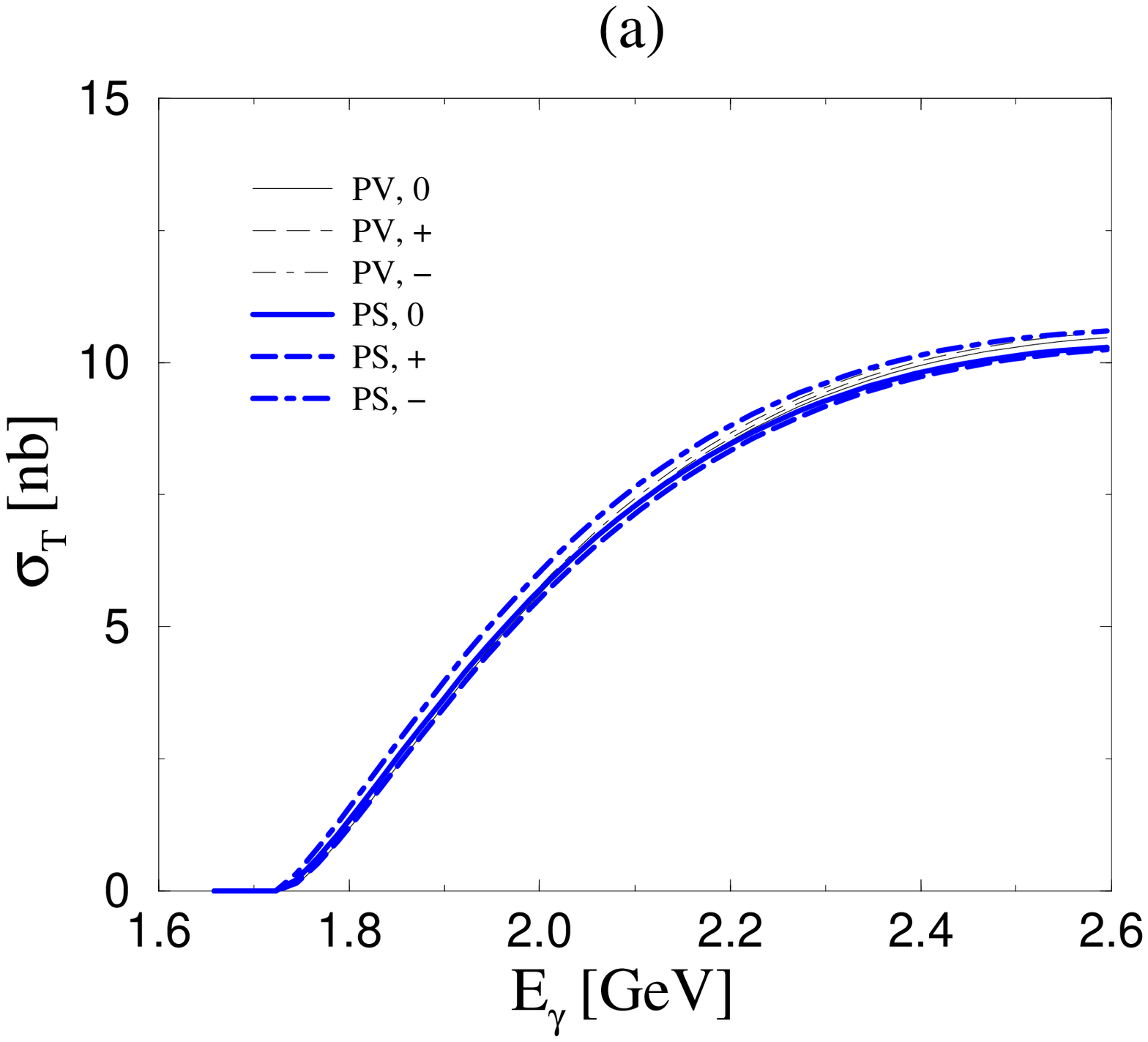}}
\resizebox{5.5cm}{4cm}{\includegraphics{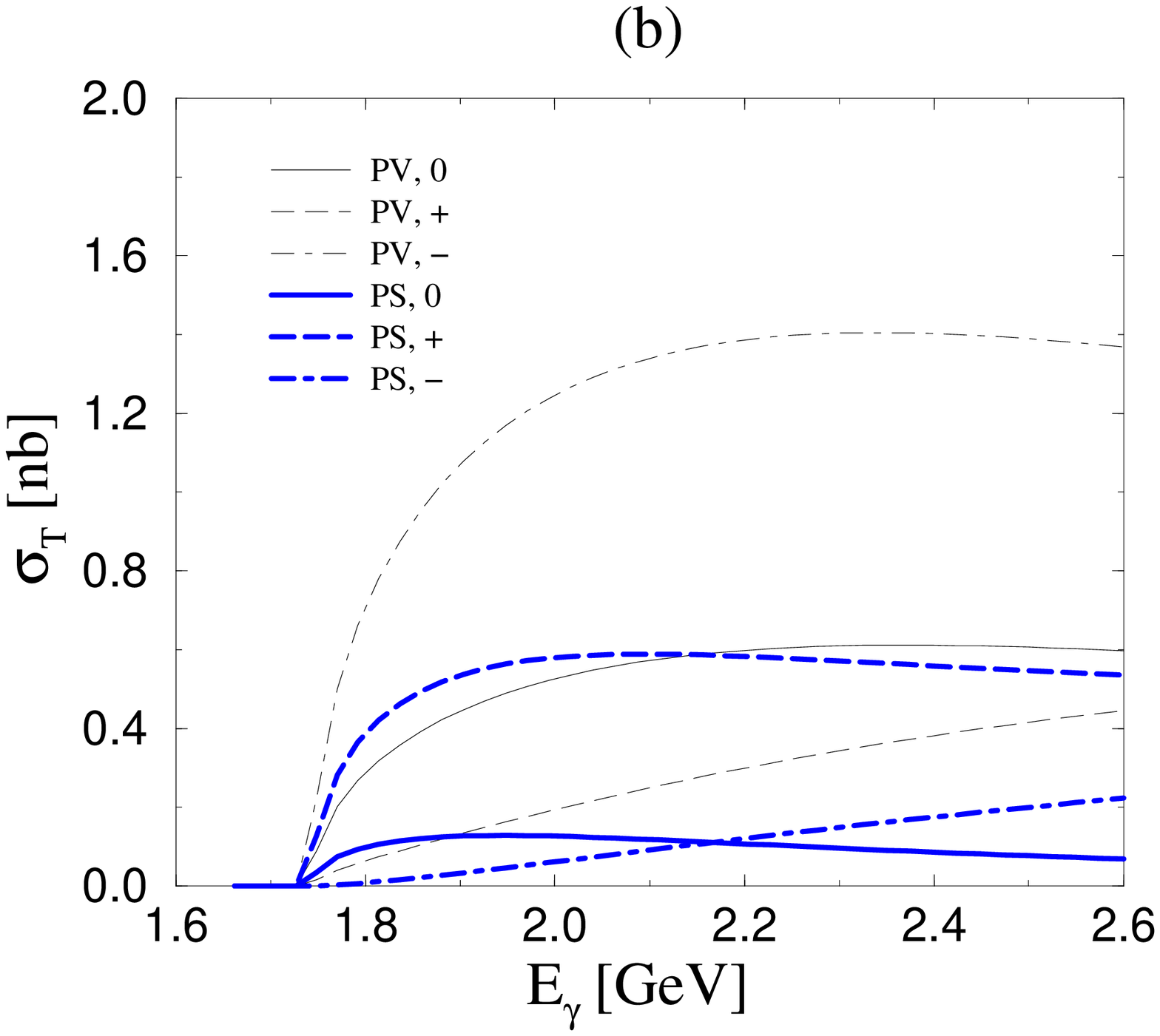}}
\end{tabular}
\caption{The total cross sections for the reactions of $\gamma
n\rightarrow K^{-}\Theta^+_{-}$ (a) and 
$\gamma p\rightarrow \bar{K}^{0}\Theta^+_{-}$ (b).}
\label{gn4}
\end{figure} 

We now present the total cross sections for the negative parity 
$\Theta^{+}_{-}$ in Fig.~\ref{gn4}.  
The contribution of $K^*$ exchange is almost
negligible in the case of the $\gamma n\rightarrow K^-\Theta^+$
process, whereas it plays a main role in 
$\gamma p\rightarrow \bar{K}^0\Theta^+$.  The total cross
sections for the negative-parity $\Theta^+$ turn out to be
approximately ten times smaller than those for the positive-parity
one.  This fact pervades rather universally in various reactions for
the $\Theta^+$ production.  The reason is that the momentum-dependent
$p$-wave coupling $\vec \sigma \cdot \vec q$ for the positive parity
$\Theta^+$ enhances the coupling strength effectively at the momentum
transfer $|\vec q| \sim 1$ GeV, a typical value for the $\Theta^+$
production using non-strange particles.  The enhancement factor is
about 1 GeV/0.26 GeV, where 0.26 GeV is the kaon momentum in the  
$\Theta^+$ decay.  Therefore, the cross sections become larger for the
positive parity case than for the negative parity case by a factor 
$(1/0.26)^2 \sim 10$.  

\begin{figure}[tbh]
\begin{tabular}{cc}
\resizebox{5.5cm}{4cm}{\includegraphics{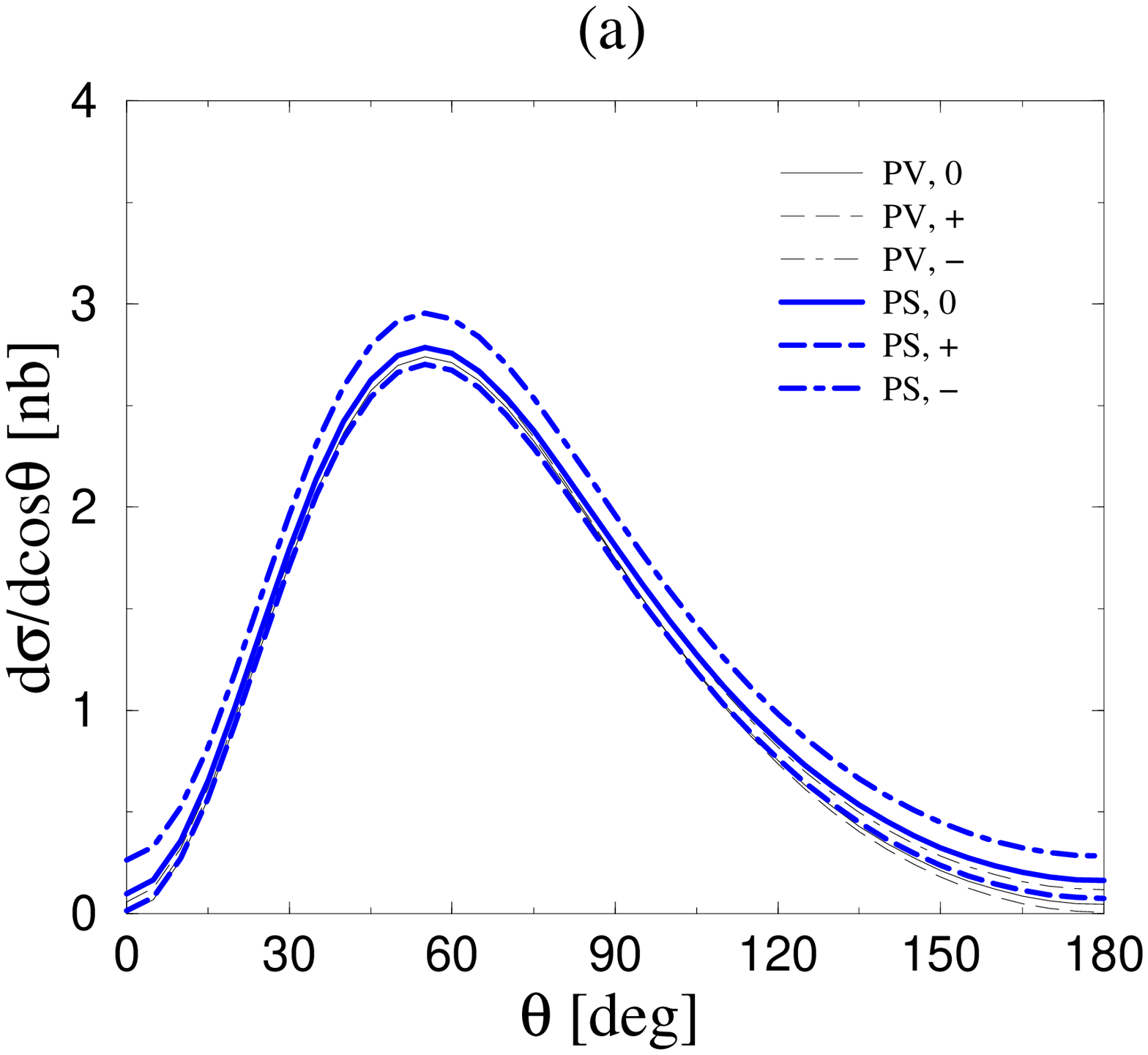}}
\resizebox{5.5cm}{4cm}{\includegraphics{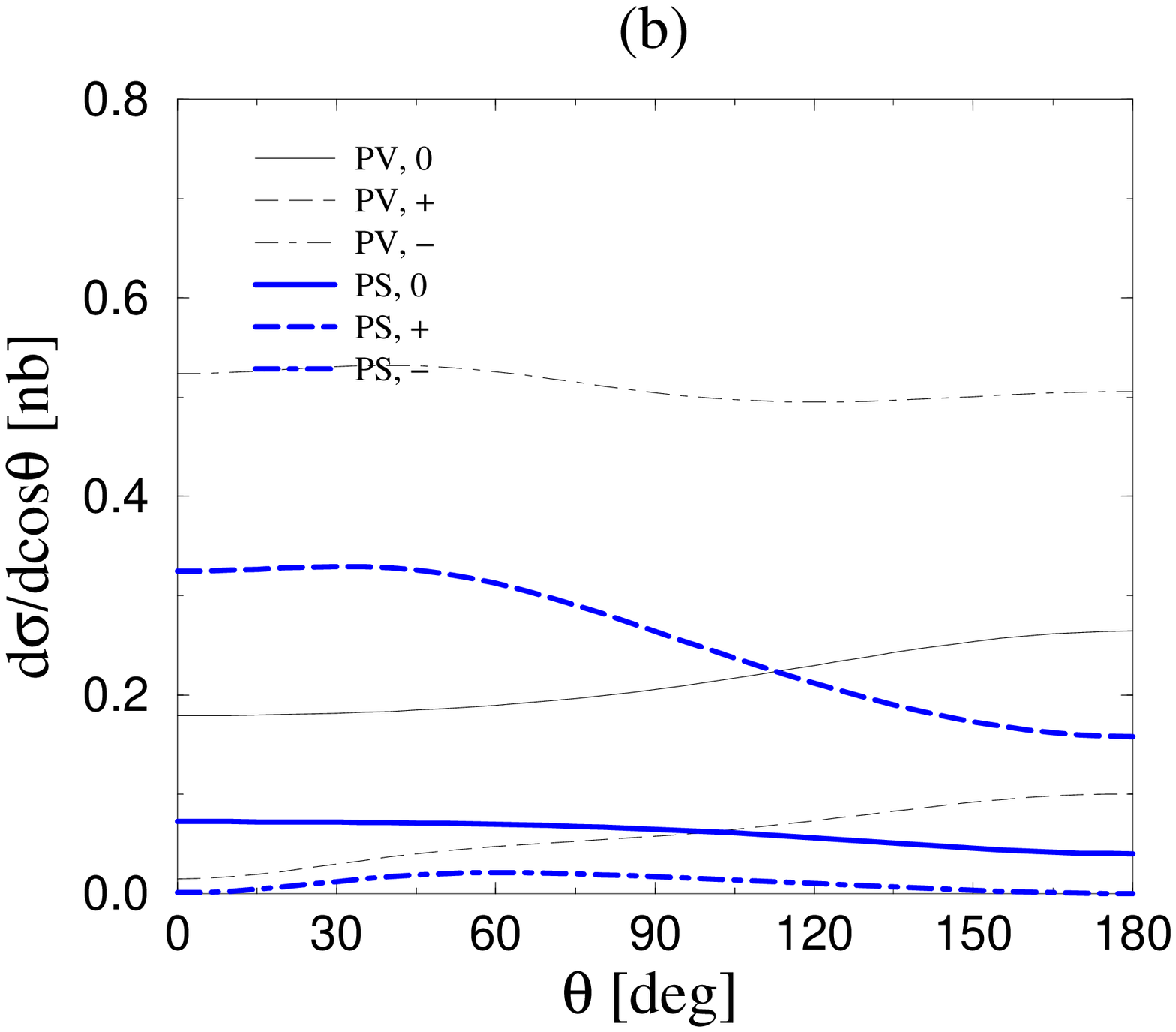}}
\end{tabular}
\caption{The differential cross sections for the reactions of $\gamma
n\rightarrow K^{-}\Theta^+_{-}$ (a) and 
$\gamma p\rightarrow \bar{K}^{0}\Theta^+_{-}$ (b) at $\sqrt{s} = 2.1\,{\rm  GeV}$.} 
\label{gn5}
\end{figure} 

The differential cross sections for the $\Theta^{+}_{-}$ photo-production are drawn in
Fig.~\ref{gn5}.  The peak around $60^\circ$ appears in the $\gamma
n$ interaction as in the case of the $\Theta^{+}_{+}$.  That for the
production via the $\gamma p$ interaction shows quite different from
the case of the $\Theta_{+}^{+}$.  
\subsection{$N N$ scattering}
\begin{figure}[tbh]
\begin{center}
\resizebox{5.5cm}{4cm}{\includegraphics{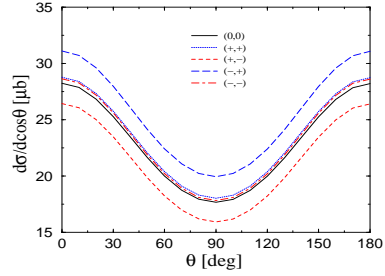}}
\caption{The differential cross sections for the reaction 
$np\rightarrow\Lambda\Theta^{+}_{+}$  at $E_{\rm CM}$ = 2.7 GeV with
five different combinations of the signs of the $K^{*}N\Theta$
coupling constants as labelled by (sgn($g_{K^*N\Theta}$), 
sgn($g_{K^*N\Theta}^T)$).  The parameter
set of the Nijmegen potential with the cutoff parameter $\Lambda$ =
1.0 GeV is employed.}  
\label{nn2}
\end{center}
\end{figure}

In this section, we present the total and differential cross sections 
for the reactions $np\rightarrow \Lambda^0 \Theta^+$ and
$np\rightarrow \Sigma^0 \Theta^+$ with two different parities of
$\Theta^{+}$.  We first consider the case of parameter set of the
Nijmegen potential.   
\begin{figure}[tbh]
\begin{tabular}{cc}
\resizebox{5.5cm}{4cm}{\includegraphics{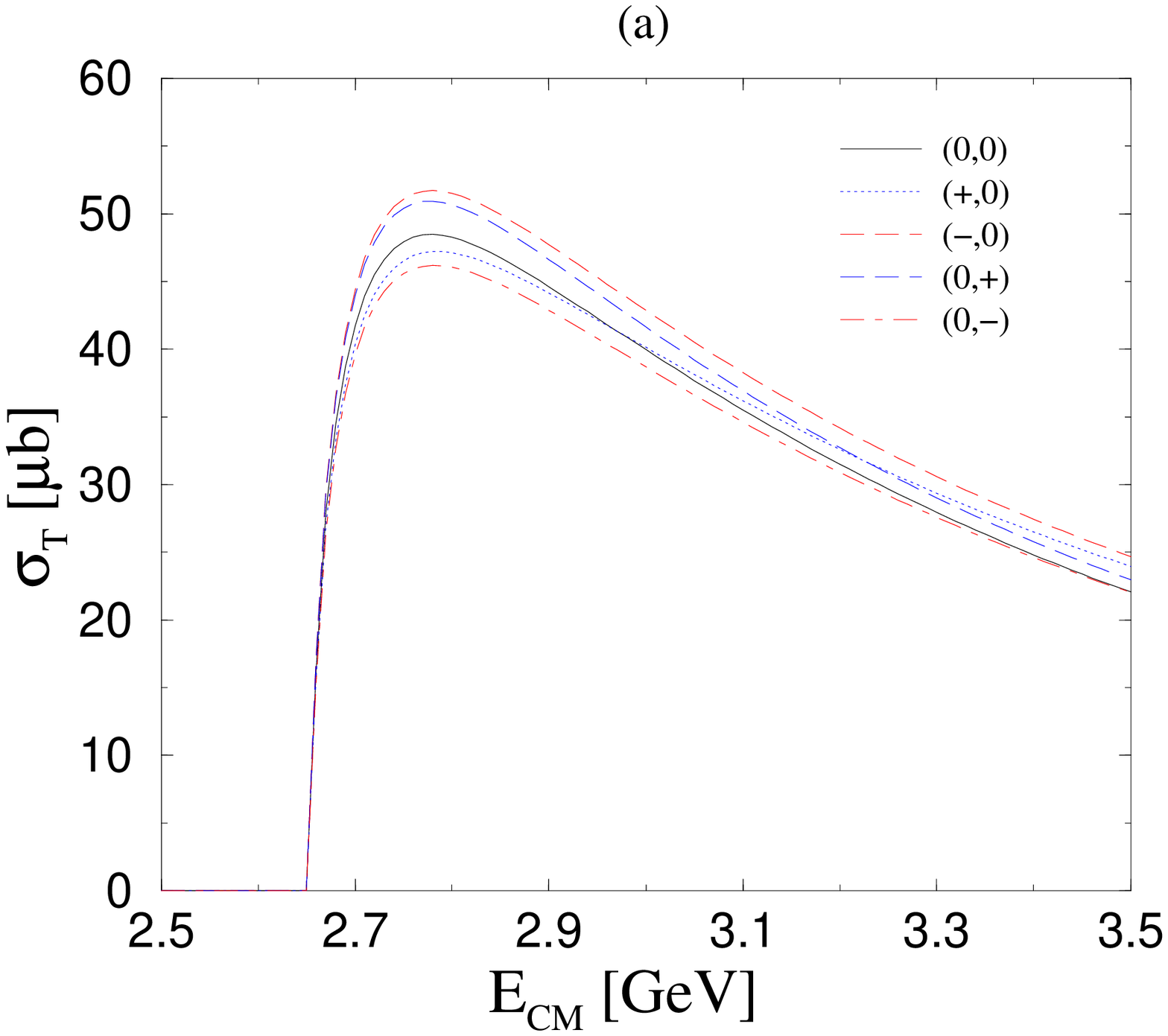}}
\resizebox{5.5cm}{4cm}{\includegraphics{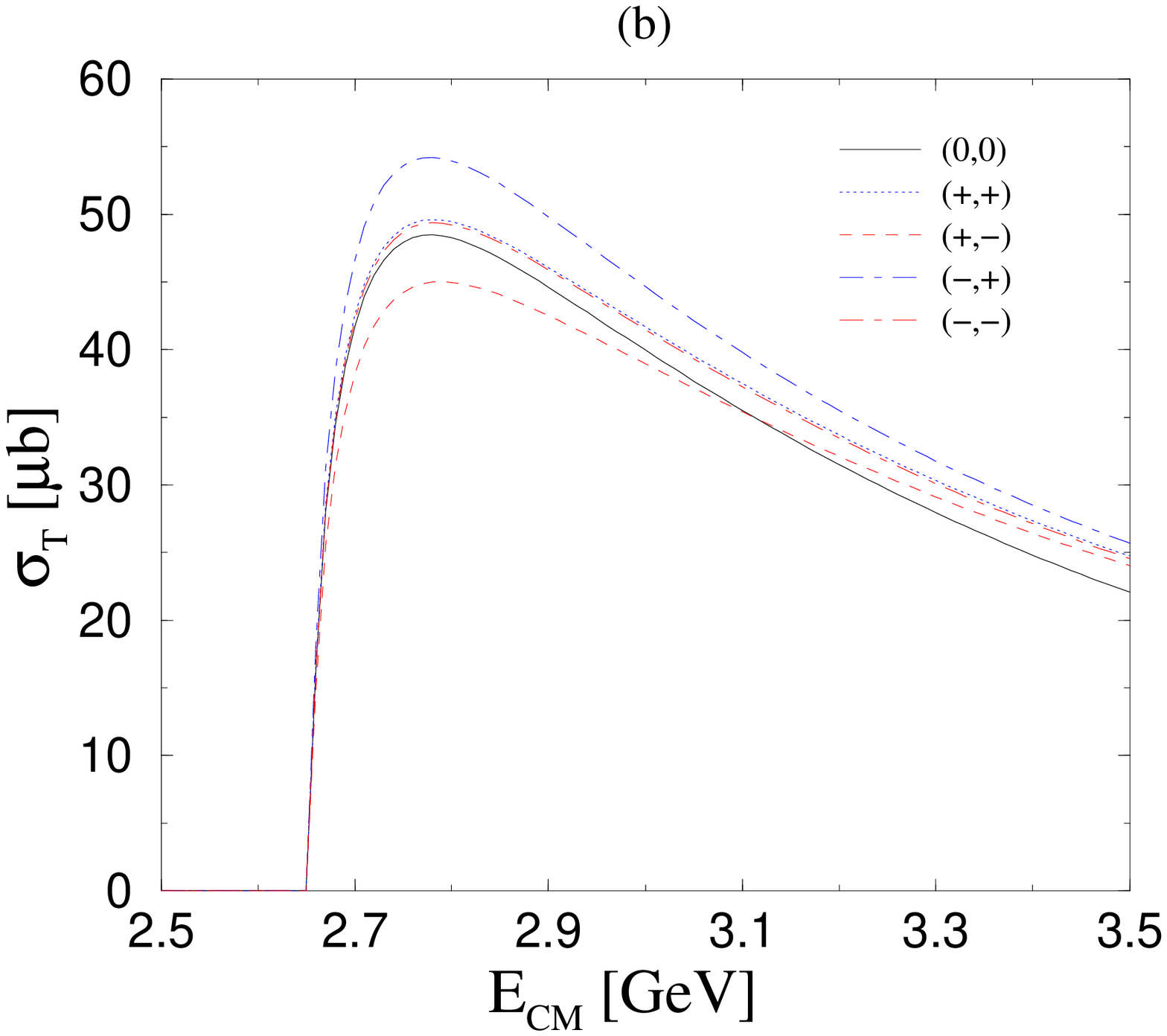}}
\end{tabular}
\caption{The total cross sections of
$np\rightarrow\Lambda\Theta^{+}_{+}$ with ten different
combinations of the signs of the $K^{*}N\Theta$ coupling constants
which are labelled 
by (sgn($g_{K^*N\Theta}$), sgn($g_{K^*N\Theta}^T)$).  The parameter
set of the Nijmegen potential with the cutoff parameter $\Lambda$ =
1.0 GeV is employed.} 
\label{nn1}
\end{figure}   
In Fig.~\ref{nn1}, we draw the total cross sections of
$np\rightarrow\Lambda\Theta^{+}_{+}$ for different signs of the
coupling constants, which are labelled as (sgn($g_{K^*N\Theta}$),
sgn($g_{K^*N\Theta}^T)$).  We compare the results from ten different
combinations of the signs.  As shown in Fig.~\ref{nn1}, the
dependence on the signs is rather weak.  Moreover, we find
that the contribution from $K^*$ exchange is very tiny.  The average
total cross section is obtained as 
$\sigma_{np\rightarrow\Lambda\Theta^{+}_{+}}\sim$  40 $\mu b$ 
in the range of the center-of-mass (CM) 
energy $E_{\rm CM}^{\rm th} \le E_{\rm CM} \le 
3.5$ GeV, where $E_{\rm CM}^{\rm th}$ = 2656 MeV. 
Since the angular distribution for all reactions is with a similar 
shape, we show the results only for the case of
$np\rightarrow\Lambda\Theta^{+}_{+}$ in Fig.~\ref{nn2}. 

\begin{figure}[tbh]
\begin{center}
\resizebox{5.5cm}{4cm}{\includegraphics{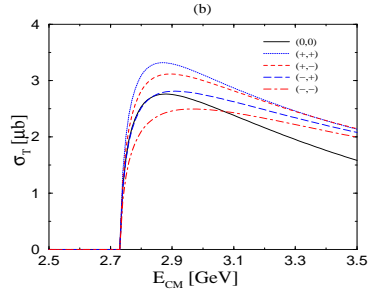}}
\caption{The total cross sections for the reaction
$np\rightarrow\Sigma^{0}\Theta^{+}_{+}$.  The parameter
set of the Nijmegen potential with the cutoff parameter $\Lambda$ = 
1.0 GeV is employed.  The notations are the same as in
Fig.~\ref{nn2}.}   
\label{nn3}
\end{center}
\end{figure}    

In Fig.~\ref{nn3}, we draw the total cross sections for the
reaction $np\rightarrow\Sigma^{0}\Theta^{+}_{+}$.  We find that they
are about ten times smaller than those for the reaction 
$np\rightarrow\Lambda\Theta^{+}_{+}$.  The corresponding average
total cross section is found to be     
$\sigma_{np\rightarrow\Sigma^0\Theta^{+}_{+}}\sim$  2.0 $\mu b$ in the
range of the CM energy $E_{\rm CM}^{\rm th} \le E_{\rm CM} \le  
3.5$ GeV, where $E_{\rm CM}^{\rm th}$ = 2733 MeV. It can be easily
understood from the fact that the ratio of the coupling constants
$|g_{KN\Lambda}/g_{KN\Sigma}| = 3.74$ is rather large and   
the contribution from $K$ exchange is dominant.  
\begin{figure}[tbh]
\begin{tabular}{cc}
\resizebox{5.5cm}{4cm}{\includegraphics{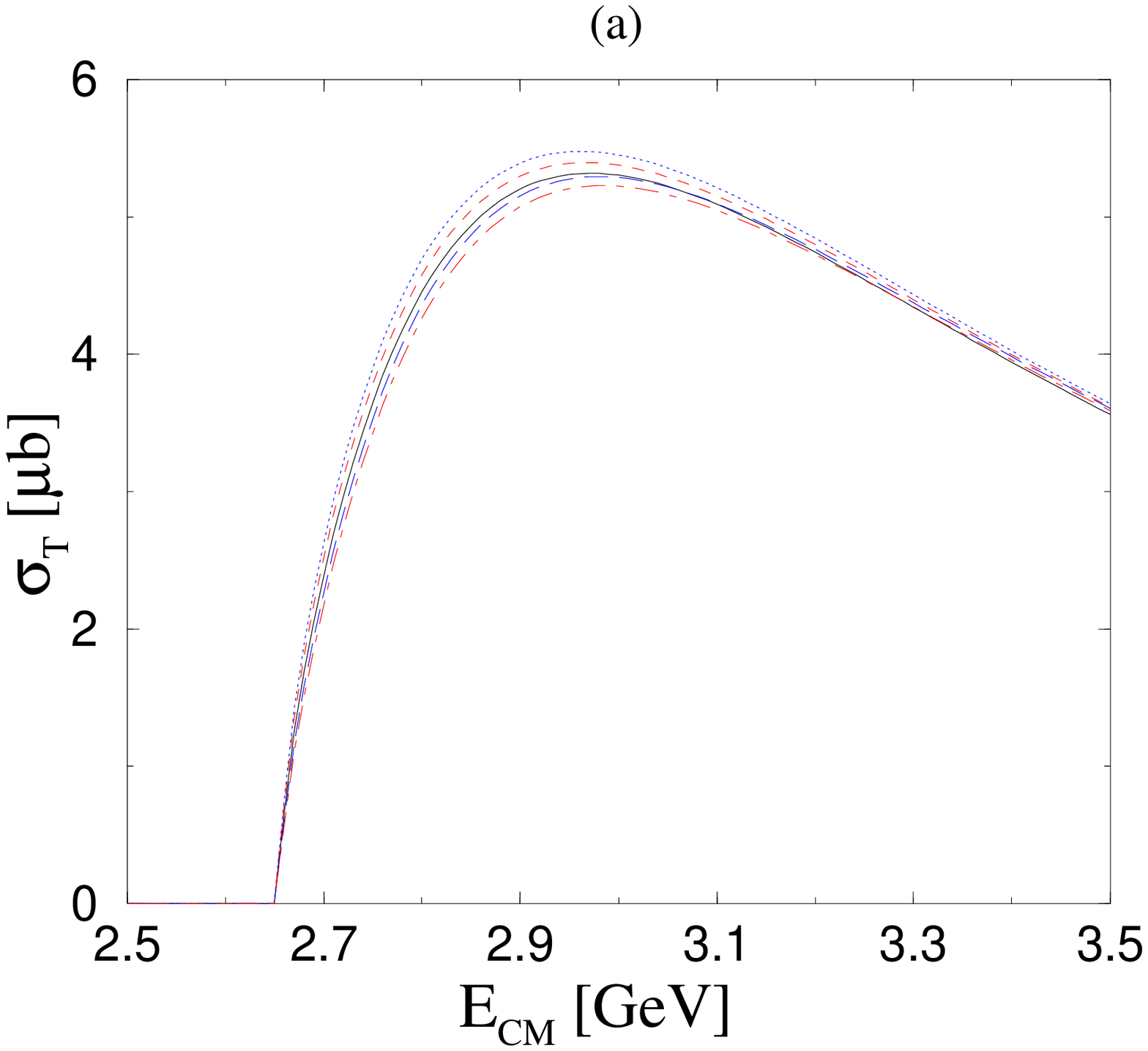}}
\resizebox{5.5cm}{4cm}{\includegraphics{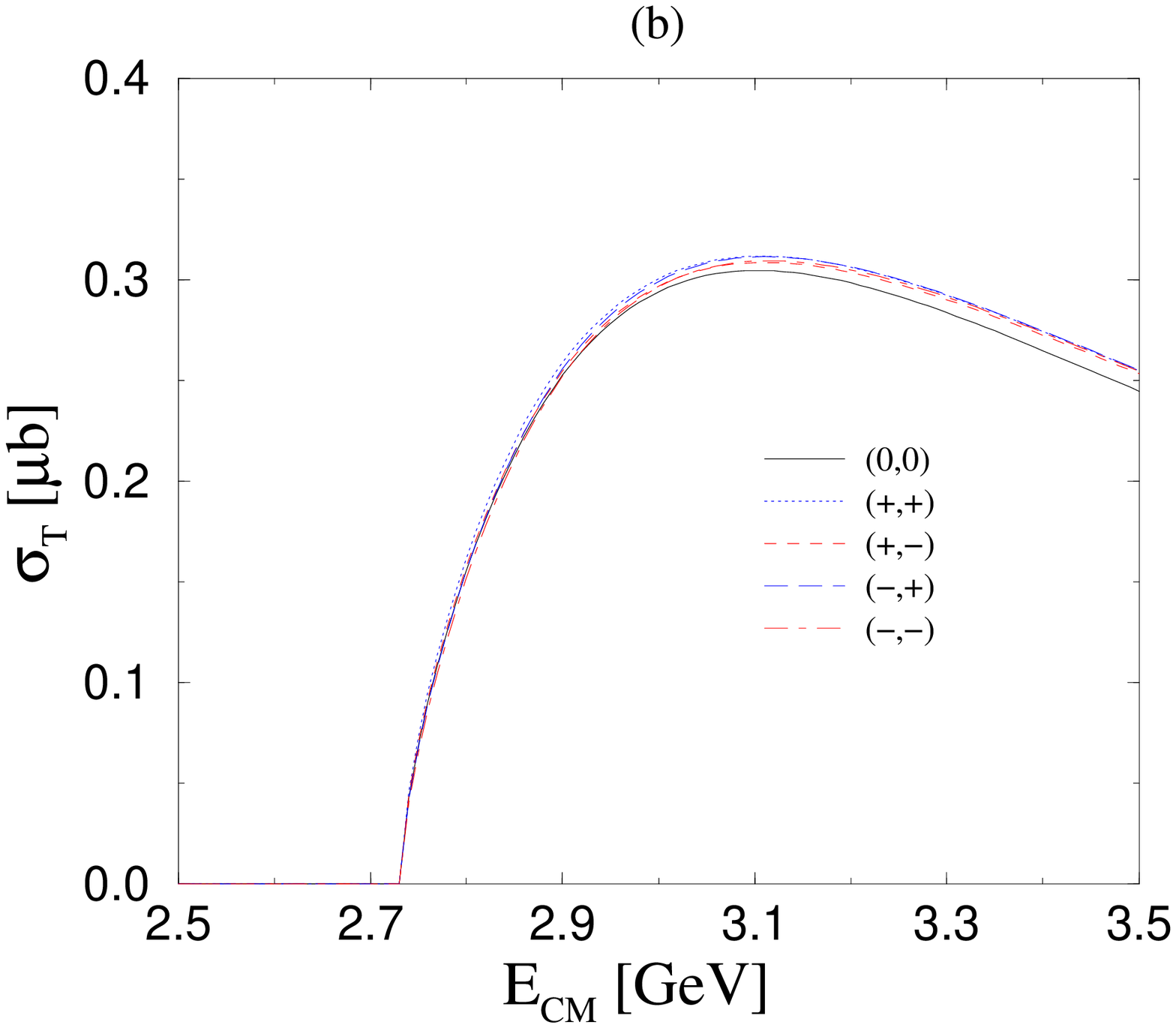}}
\end{tabular}
\caption{The total cross sections of
$np\rightarrow\Lambda\Theta^{+}_{-}$ in the left panel (a) and
$np\rightarrow\Sigma^{0}\Theta^{+}_{-}$ in the right panel (b).  The
parameter set of the Nijmegen potential with the cutoff parameter
$\Lambda$ = 1.0 GeV is employed.  The notations are the same as in
Fig.~\ref{nn2}.}   
\label{nn3}
\end{figure}   

As for the negative parity $\Theta^{+}$, we show the results in
Fig.~\ref{nn3}.  Once again we find that the contribution of
$K^{*}$ exchange plays only a minor role.  We observe in average that
$\sigma_{np\rightarrow\Lambda\Theta^{+}_{-}}\sim$  5.0 $\mu b$ and   
$\sigma_{np\rightarrow\Sigma^{0}\Theta^{+}_{-}}\sim$  0.3 $\mu b$ in the
range of the CM energy $E_{\rm CM}^{\rm th} \le E_{\rm CM} \le  
3.5$ GeV. They are almost ten times smaller than those of
$\Theta^{+}_{+}$.  This behavior can be interpreted dynamically
by the fact that a large momentum transfer $\sim 800\, {\rm MeV}$
enhances the P-wave coupling of the $\Theta^{+}_{+}$ than the S-wave one
of the $\Theta^{+}_{-}$.

\begin{figure}[tbh]
\begin{tabular}{cc}
\resizebox{5.5cm}{4cm}{\includegraphics{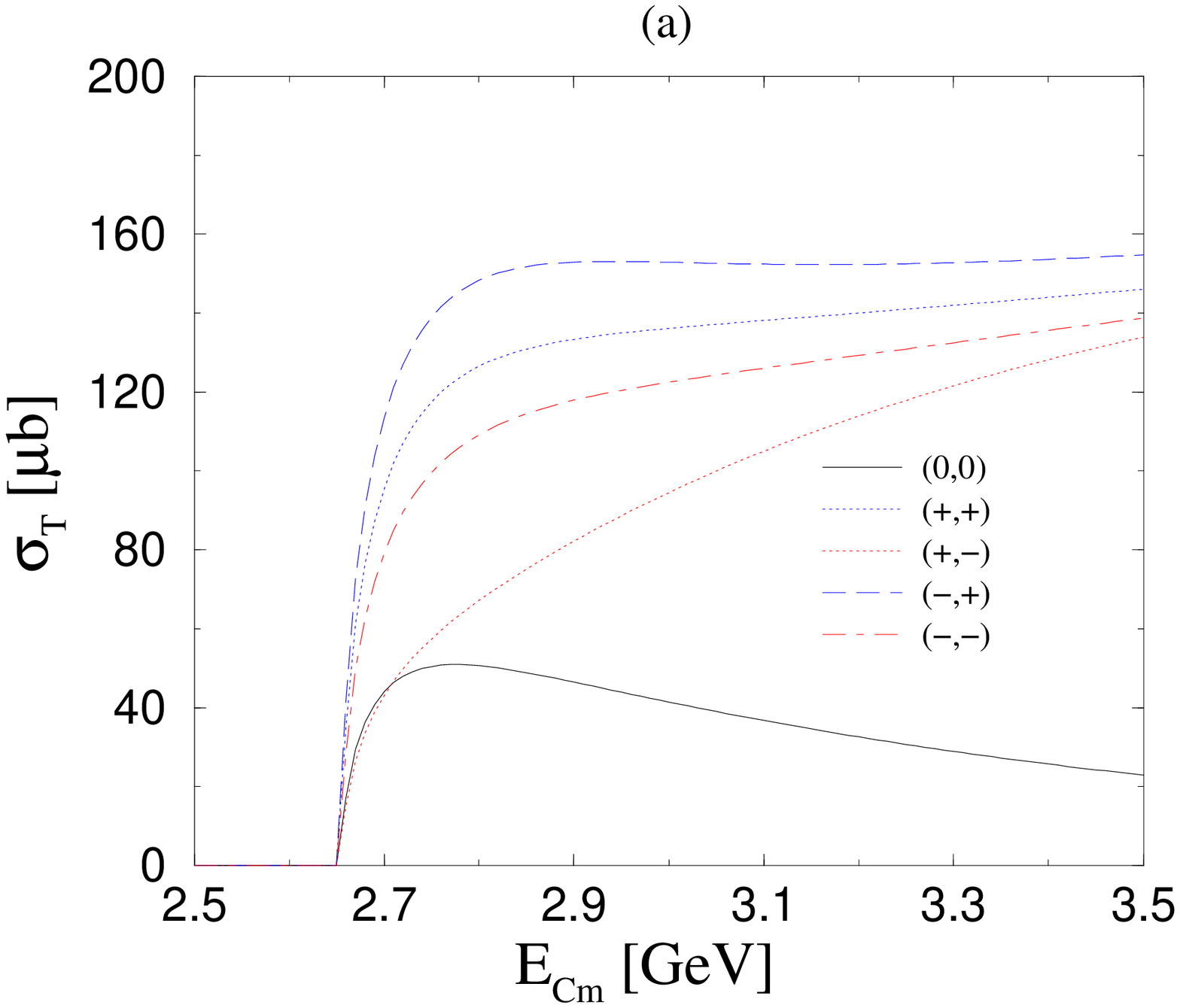}}
\resizebox{5.5cm}{4cm}{\includegraphics{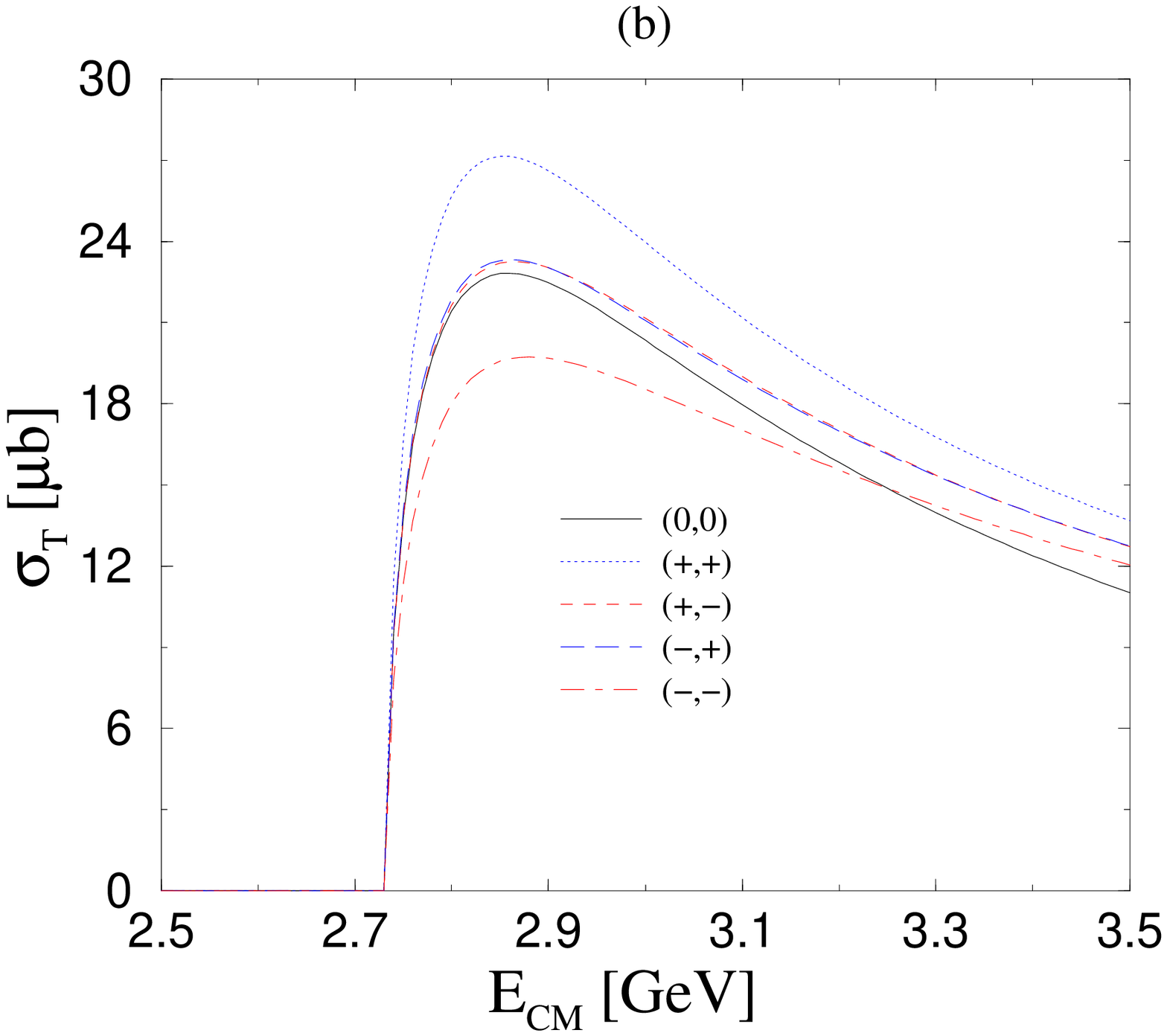}}
\end{tabular}
\caption{The total cross sections of
$np\rightarrow\Lambda\Theta^{+}_{+}$ in the left panel (a) and
$np\rightarrow\Sigma^{0}\Theta^{+}_{+}$ in the right panel (b).  The
parameter set of the J\"ulich--Bonn potential is employed.  The
notations are the same as in Fig.~\ref{nn2}.}
\label{nn4}
\end{figure}

In Fig.~\ref{nn4}, we show the total cross sections of the
reactions for the $\Theta^{+}_{\pm}$ with the parameter set of the
J\"ulich--Bonn potential.  Here, different cutoff parameters are
employed at different vertices as mentioned previously.   
We find that the contribution from $K^*$ exchange turns out to be
larger in the $np\rightarrow\Lambda\Theta^{+}_{+}$ reaction
than in the $np\rightarrow\Sigma^{0}\Theta^{+}_{+}$.  This can be
easily understood from the fact that the J\"ulich--Bonn cutoff parameter
$\Lambda_{K^*N\Lambda}$ is chosen to be approximately twice as large as
that of the $KN\Lambda$ vertex, while the value of the
$\Lambda_{K^*N\Sigma}$ is about two times smaller than that of the
$\Lambda_{KN\Sigma}$.  The average total cross sections are obtained
as follows: $\sigma_{np\rightarrow\Lambda\Theta^{+}_{+}}\sim$  100
$\mu b$ and $\sigma_{np\rightarrow\Sigma^{0}\Theta^{+}_{+}}\sim$  20
$\mu b$ in the range of the CM energy 
$E_{\rm CM}^{\rm th} \le E_{\rm CM} \le 3.5$ GeV.
  
\begin{figure}[tbh]
\begin{tabular}{cc}
\resizebox{5.5cm}{4cm}{\includegraphics{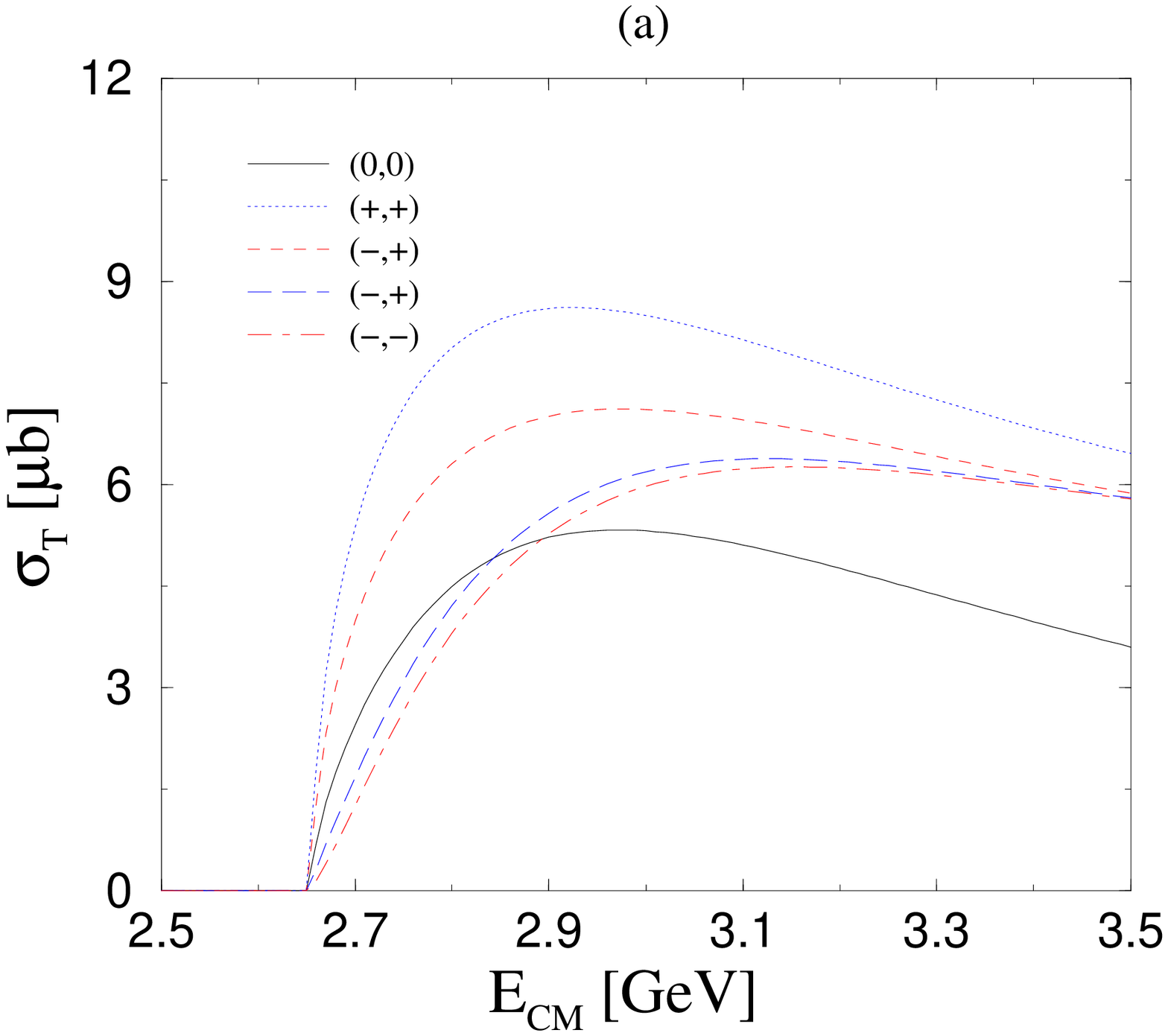}}
\resizebox{5.5cm}{4cm}{\includegraphics{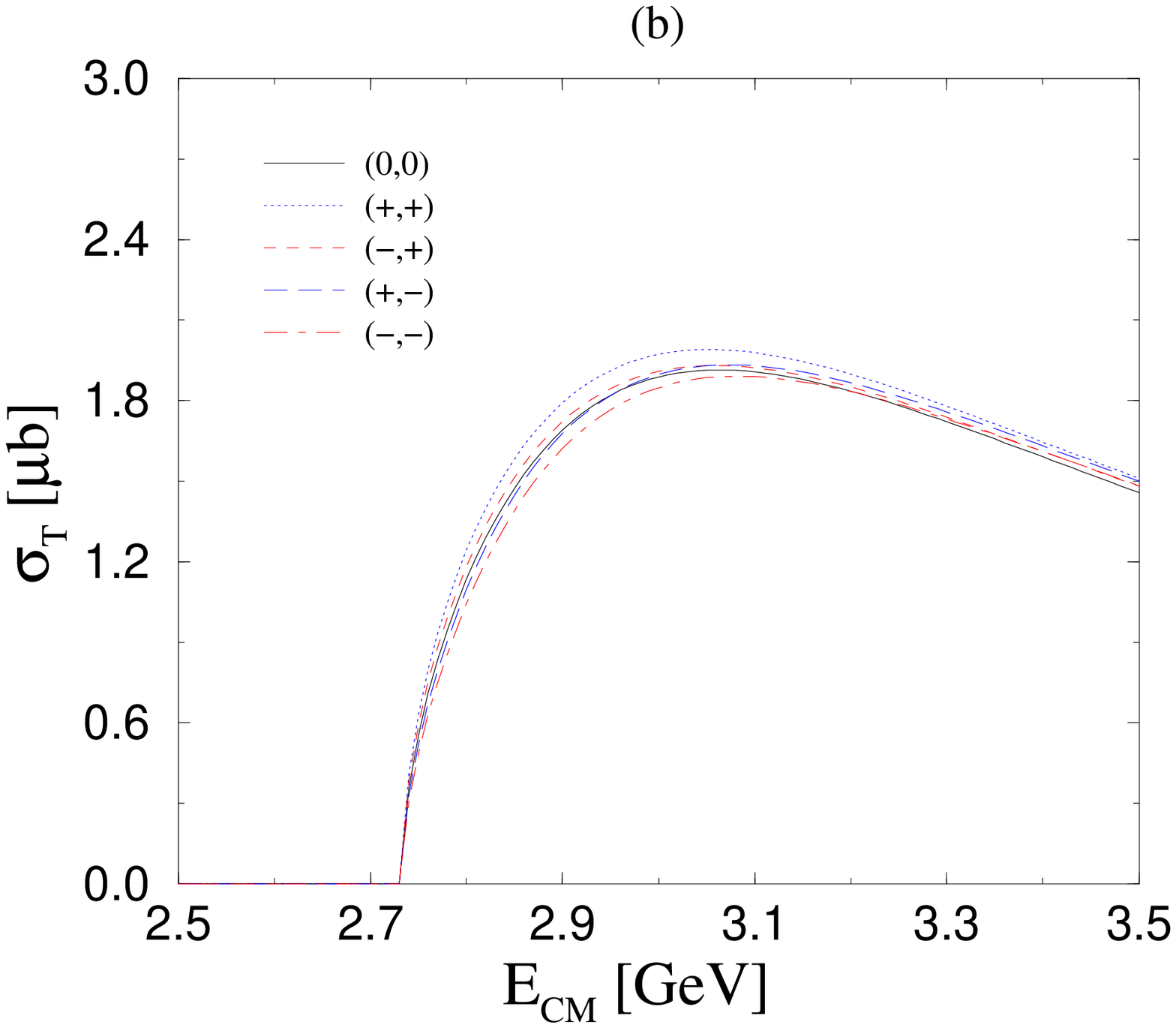}}
\end{tabular}
\caption{The total cross sections of
$np\rightarrow\Lambda\Theta^{+}_{-}$ in the left panel (a) and
$np\rightarrow\Sigma^{0}\Theta^{+}_{-}$ in the right panel (b).  The
parameter set of the J\"ulich-Bonn potential is employed.  The
notations are the same as in Fig.~\ref{nn2}.}
\label{nn5}
\end{figure}   
In Fig.~\ref{nn5}, the total cross sections for $\Theta^{+}_{-}$ are
drawn.  In this case, the average total cross sections are given as
follows: $\sigma_{np\rightarrow\Lambda\Theta^{+}_{+}}\sim$ 
6.0 $\mu b$ and $\sigma_{np\rightarrow\Sigma^{0}\Theta^{+}_{+}}\sim$  2.0
$\mu b$ in the 
same range of the CM energy.  The results for the negative-parity
$\Theta_{-}^{+}$ are about fifteen times smaller than those of
$\Theta^{+}_{+}$. Compared to the results with the parameter set of the Nijmegen
potential, those with the J\"ulich--Bonn one are rather sensitive to
the signs of the coupling constants.  It is due to the fact that
the cutoff parameters taken from the J\"ulich--Bonn potential are
different at each vertex.  If we had taken similar values of the
cutoff parameters for the Nijmegen potential, we would have obtained
comparable results to the case of the J\"ulich-Bonn potential.

\subsection{Polarized $pp$ scattering}
\begin{figure}[tbh]
\begin{center}
\begin{tabular}{cc}
\resizebox{5.5cm}{4cm}{\includegraphics{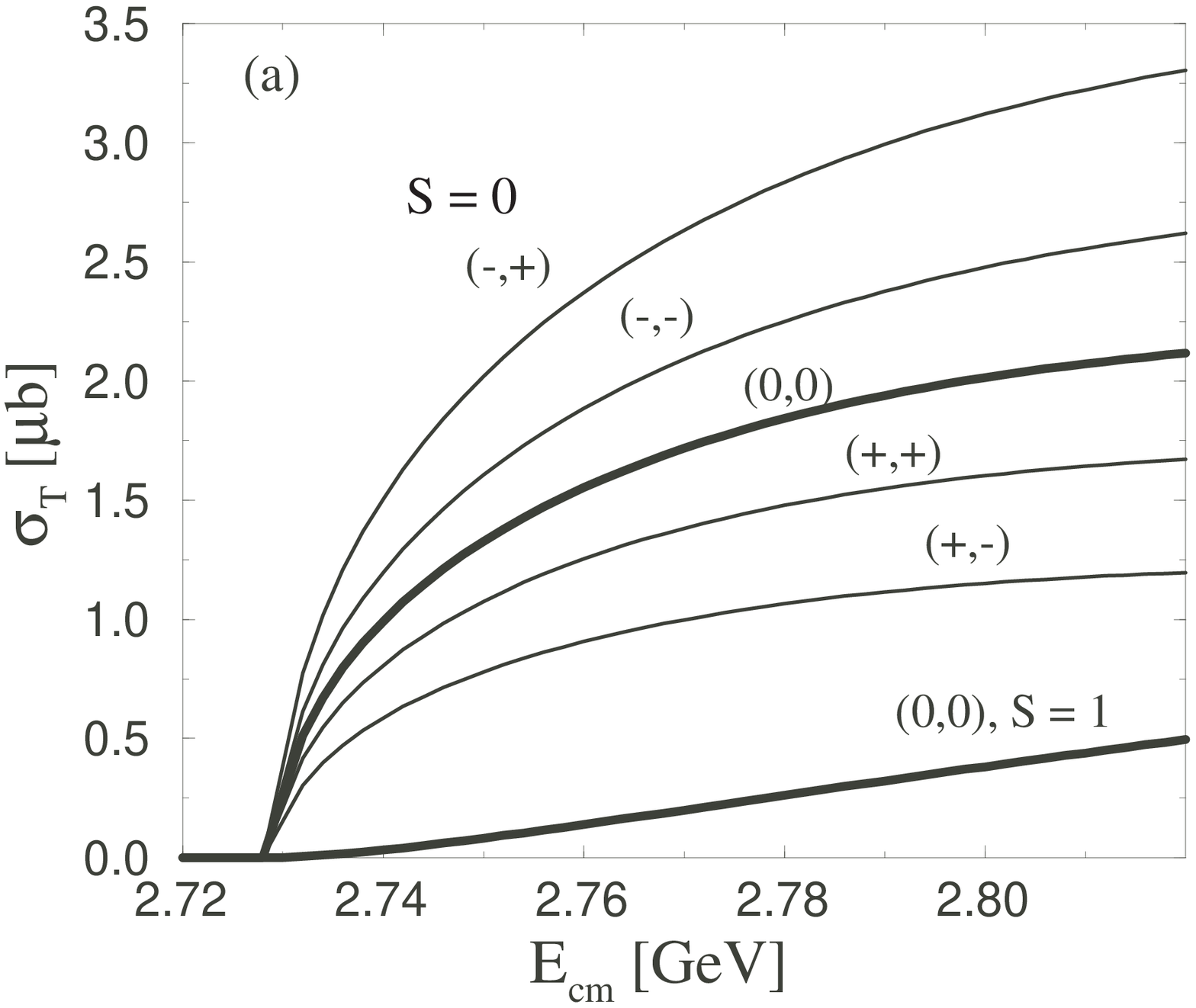}}
\resizebox{5.5cm}{4cm}{\includegraphics{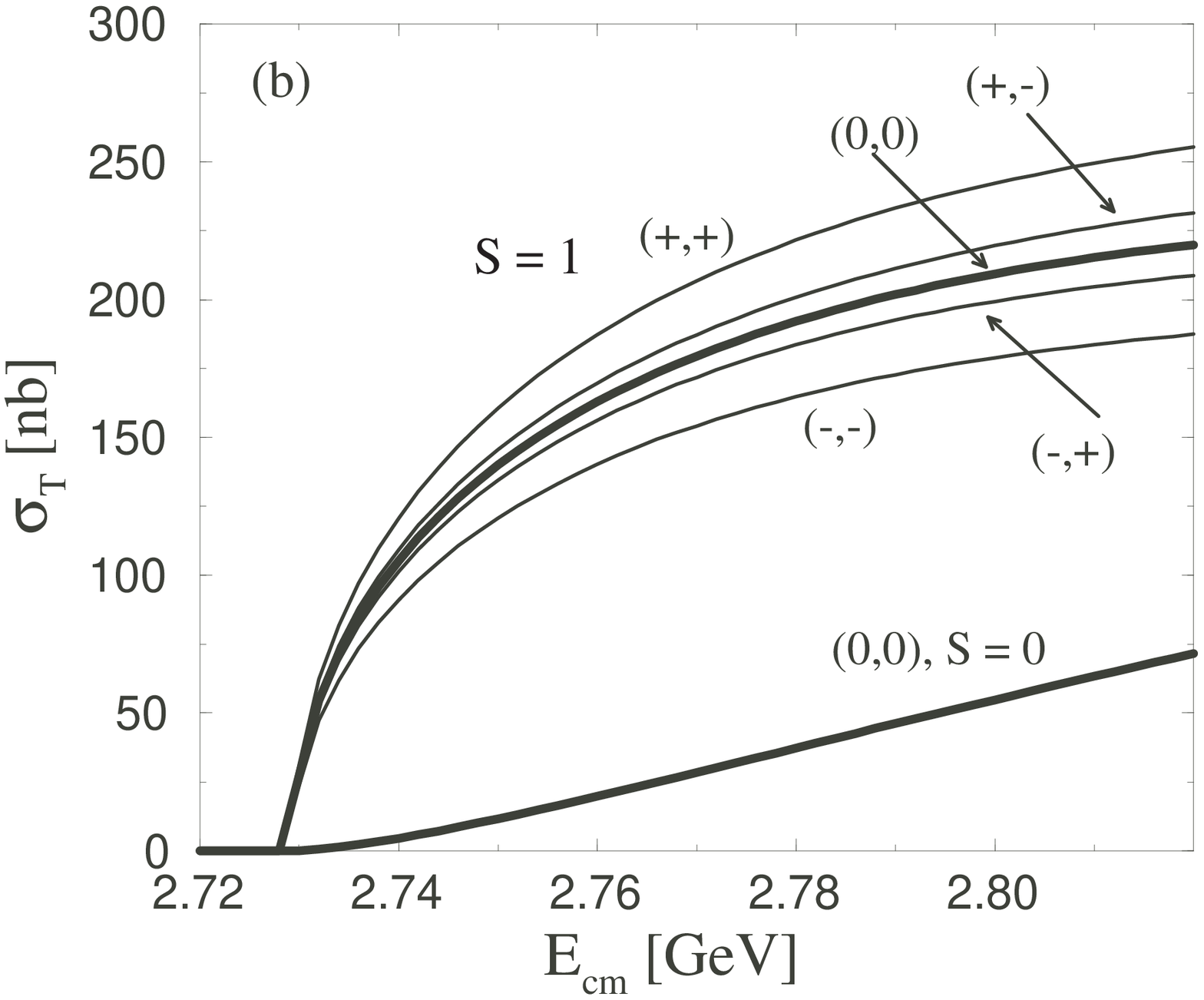}}
\end{tabular}
\caption{Total cross sections near the threshold:   
(a) for positive parity $\Theta^+$ where the allowed channel 
is ($S = 0$, even $l$) and 
(b) for negative parity $\Theta^+$ where the allowed channel 
is ($S = 1$, odd $l$).  
The labels (+,+) etc denote the signs of $g_{K^*N\Theta}$ and 
$g_{K^*N\Theta}^T$ relative to $g_{KN\Theta}$.  
The solid lines in the bottom is the cross sections 
for the forbidden channels.  
\label{stot}}
\end{center}
\end{figure}

In this section, we present the polarized $pp$ scattering results. We
note that, here, we employ only the Nijmegen potential
and the form factor of Eq.~(\ref{ff}) instead of using the J\"ulich-Bonn
potential. In Fig~\ref{stot}, total cross sections near threshold
region are   
shown as functions of the energy in the center of mass system
$\sqrt{s}$ ($\sqrt{s}_{\rm th} = 2729.4$ MeV).  
The left (right) panel is for the positive (negative) parity 
$\Theta^+$ where the allowed initial state has $S=0$ and even $l$ 
($S=1$ and odd $l$).  
For the allowed channels, five curves are shown using different 
coupling constants of $g_{K^*N\Theta}$ and $g_{K^*N\Theta}^T$; 
zero and four
different combinations of signs with the absolute values 
$|g_{K^*N\Theta}^T| = 2|g_{K^*N\Theta}| = |g_{KN\Theta}|$, as indicated 
by the pair of labels in the figures, 
(sgn($g_{K^*N\Theta}$), sgn($g_{K^*N\Theta}^T)$).  
As shown in the figure, cross sections vary with about 50 \% 
from the mean value for the vanishing $K^*$ exchanges.   
For the forbidden channels only the case of vanishing $K^*N\Theta$ 
coupling constants is shown; cross sections using finite coupling constants 
vary within about 50 \% just as for the allowed channels.  
In both figures, the s-wave threshold behavior is seen for the 
allowed channels as proportional to $(s - s_{th})^{1/2}$, while the 
forbidden channels exhibit the p-wave dependence of $(s - s_{th})^{3/2}$ 
and with much smaller values than the allowed channel. 
The suppression factor is given roughly by 
[(wave number)$\cdot$(interaction range)]$^2$ 
$\sim k/m_K \sim 0.1$ ($k = \sqrt{2m_{K}E}$), as consistent with 
the results shown in the figures. From these results, we conclude that 
the absolute value of the total cross section is 
of the order 1 [$\mu$b] for the positive parity $\Theta^+$ and 
of the order 0.1 [$\mu$b] for the negative parity $\Theta^+$.  
The fact that the positive parity case has larger cross section 
is similar to what was observed in the photoproduction and hadron
induced reaction also. This is due to the p-wave nature of the $KN\Theta$ coupling with a 
relatively large momentum transfer for the $\Theta^+$ production.  
When the smaller decay width of $\Theta^+$ is used, the result
simply scales as proportional to the width, if the 
$K^*N\Theta$ couplings are scaled similarly.  

\begin{figure}[tbh]
\begin{center}
\begin{tabular}{cc}
\resizebox{5.5cm}{4cm}{\includegraphics{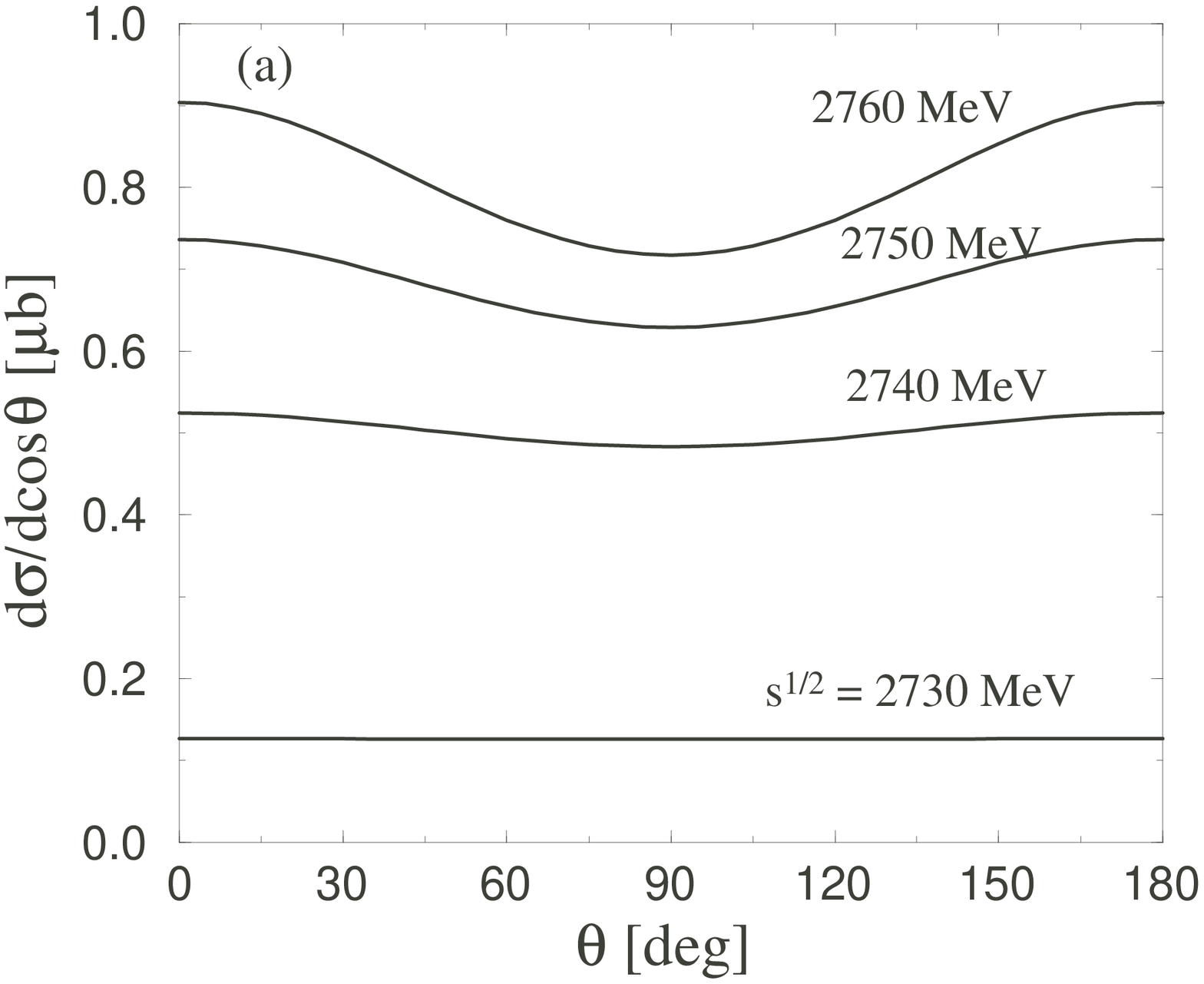}}
\resizebox{5.5cm}{4cm}{\includegraphics{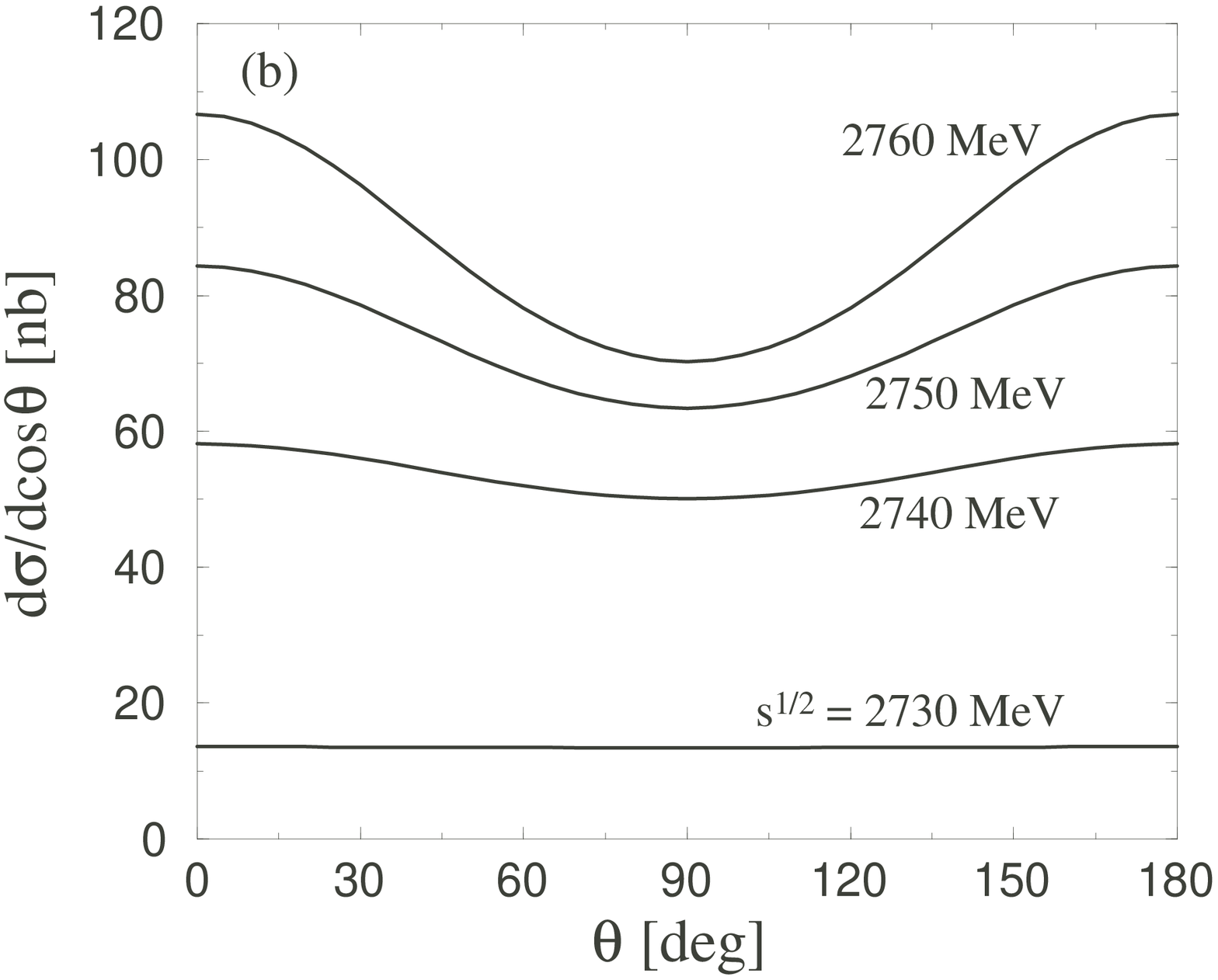}}
\end{tabular}
\caption{Angular dependence of the production cross sections 
near the threshold in the center of mass frame:  
(a) for positive parity $\Theta^+$ and 
(b) for negative parity $\Theta^+$.  
The labels denote the total incident energy $\sqrt{s}$.  
\label{dsdt}}
\end{center}
\end{figure}

\begin{figure}[tbh]
\begin{tabular}{cc}
\resizebox{5.5cm}{4cm}{\includegraphics{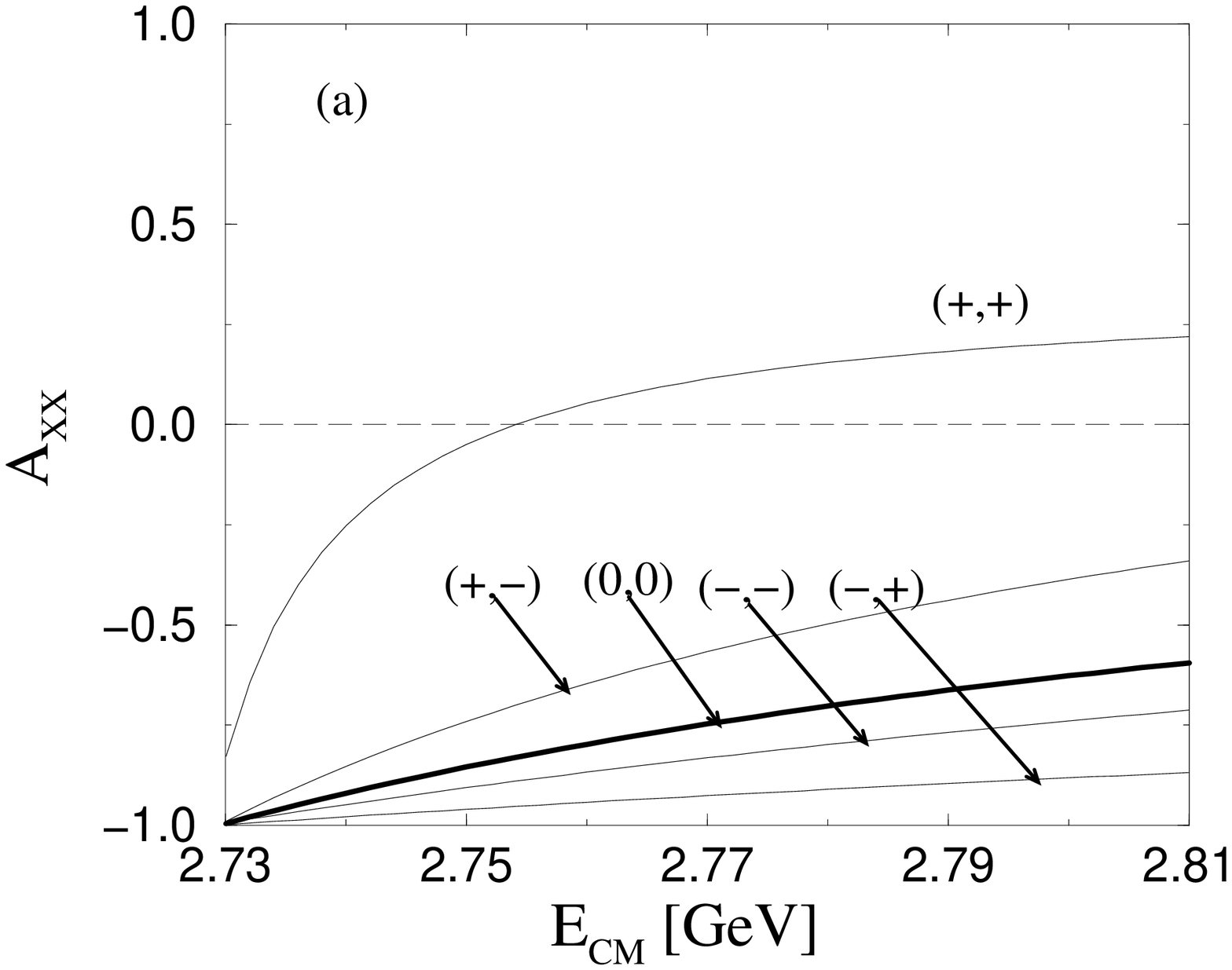}}
\resizebox{5.5cm}{4cm}{\includegraphics{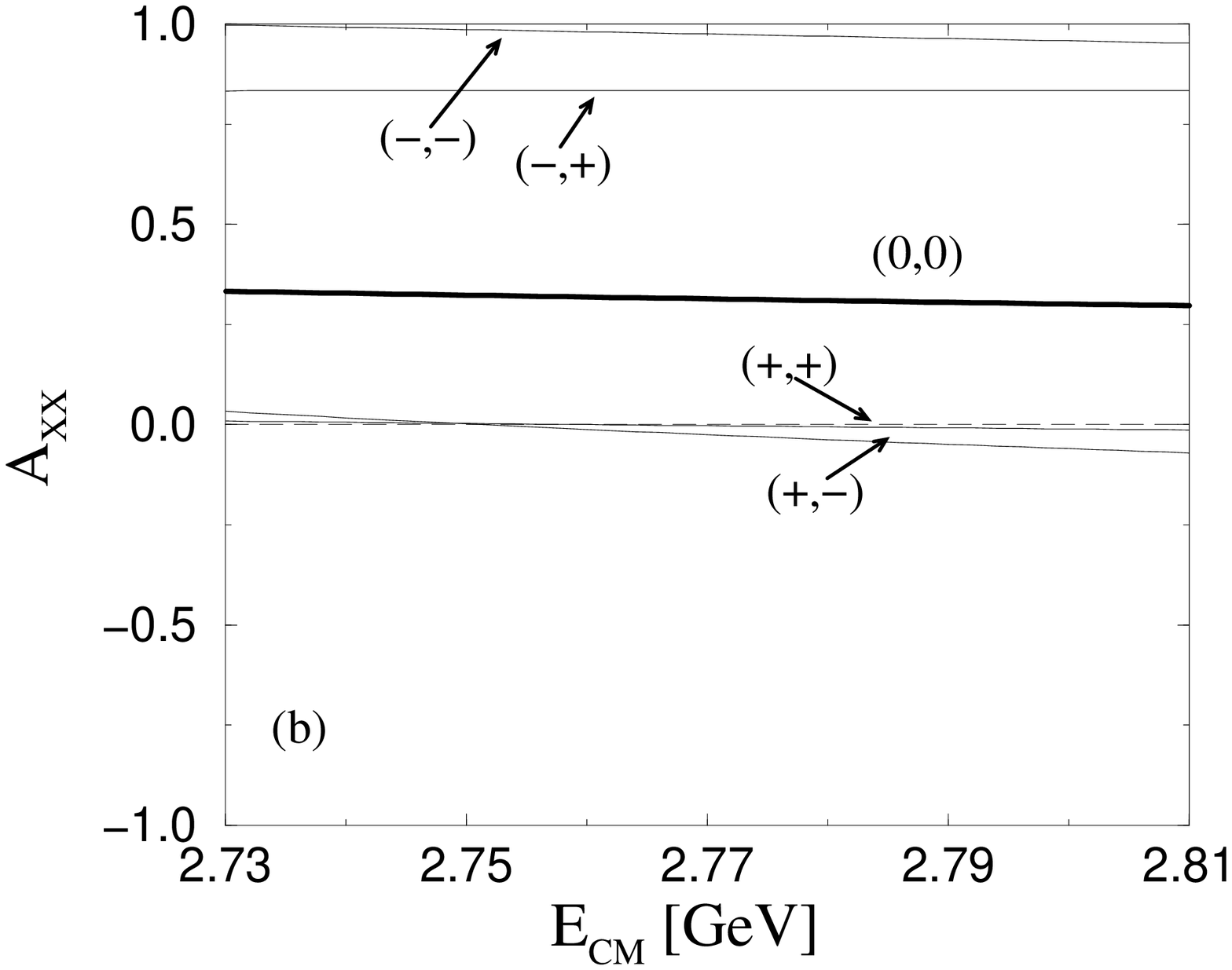}}
\end{tabular}
\begin{tabular}{cc}
\resizebox{5.5cm}{4cm}{\includegraphics{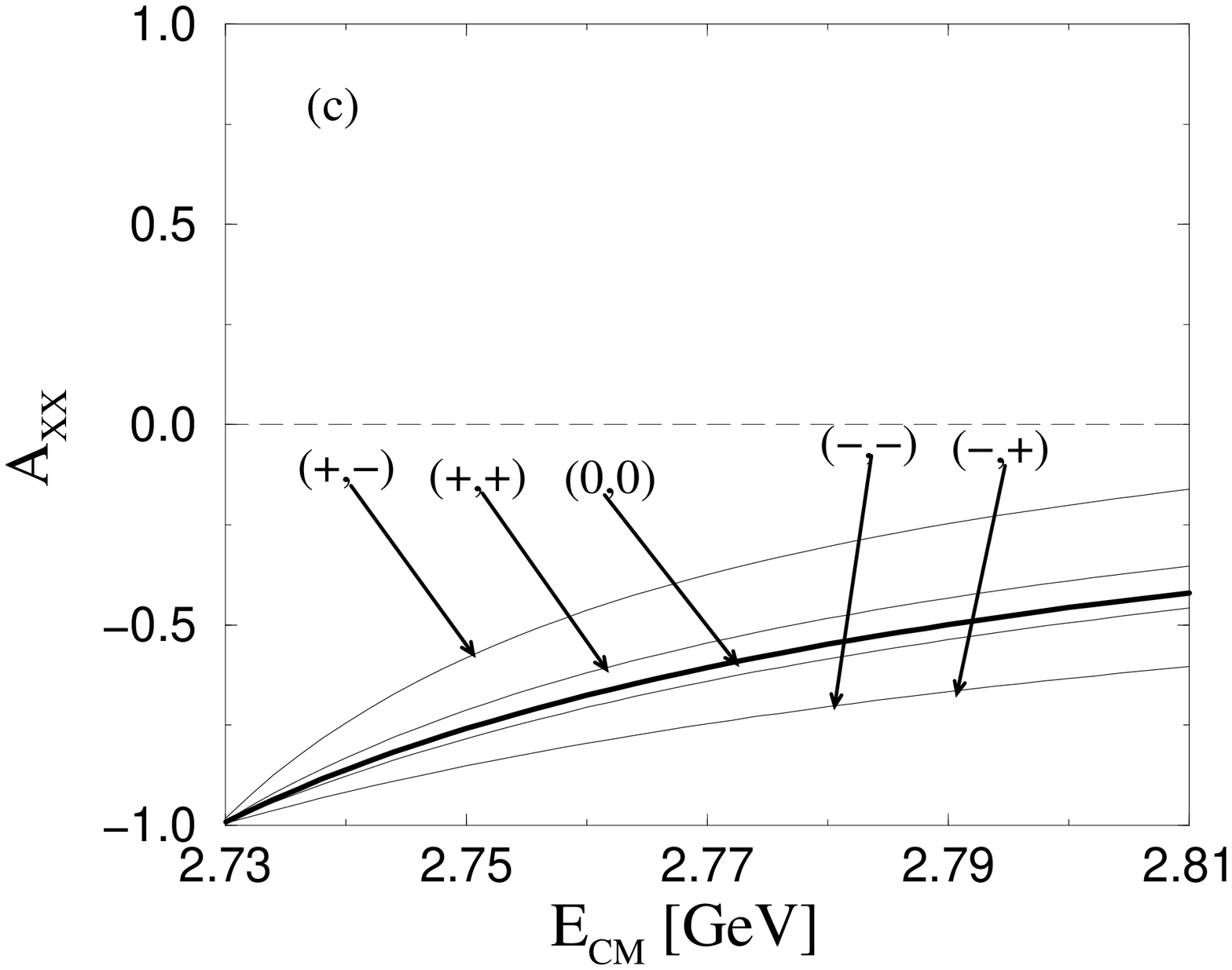}}
\resizebox{5.5cm}{4cm}{\includegraphics{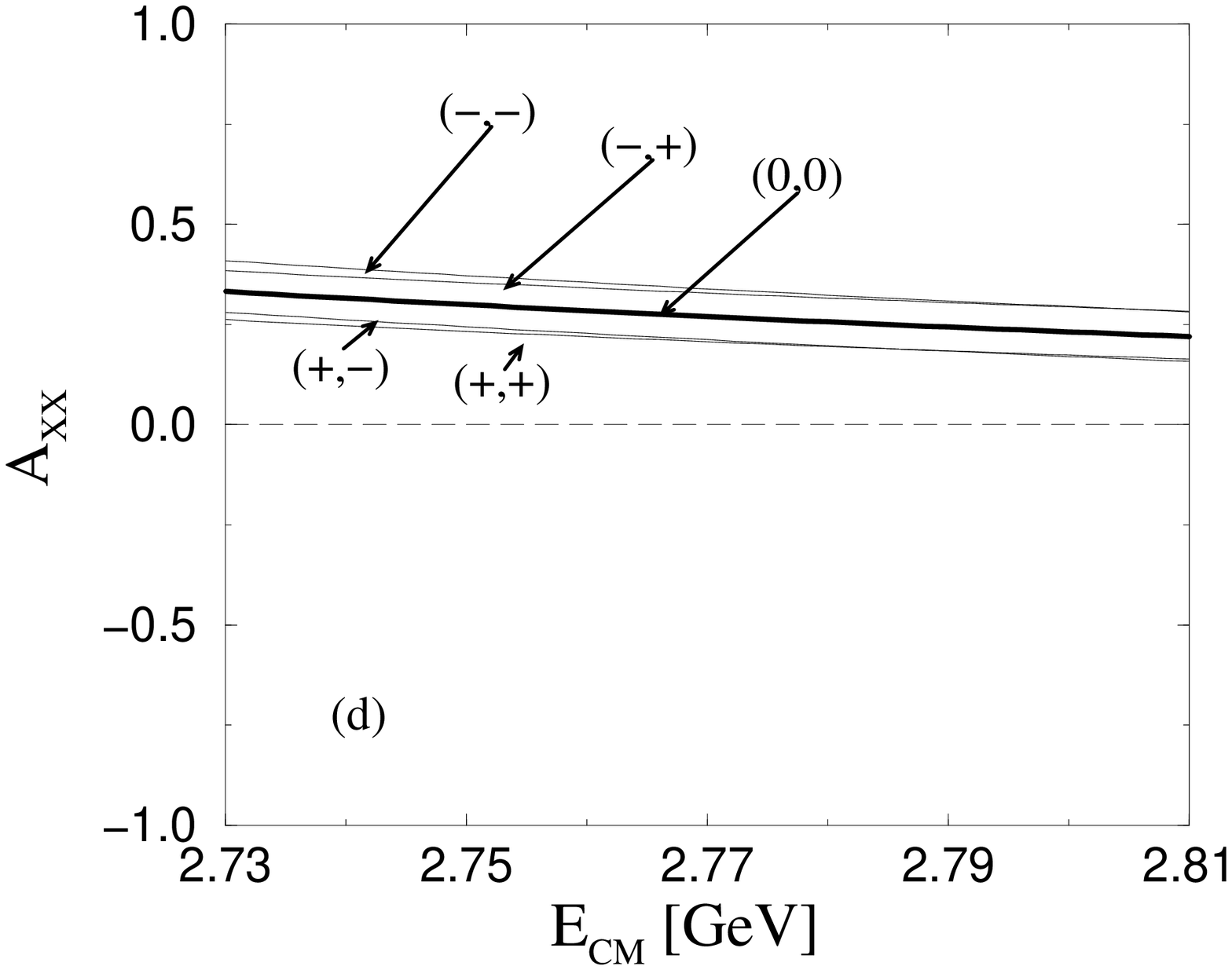}}
\end{tabular}
\caption{$A_{xx}$ for the positive (a) and negative (b) parities are
drawn without the form factor. As for the cases with the form factor, we
also show it for the positive (c) and negative (d) ones.}
\label{axx}
\end{figure}

In Fig.~\ref{dsdt}, we show the angular dependence of the cross
section in the center 
of mass system for several different energies above the threshold, 
$\sqrt{s} =$ 2730, 2740, 2750 and 2760 MeV.  
Here only $K$ exchange is included but without $K^*$ exchanges. 
The angular dependence with the $K^*$ exchanges included is similar 
but with absolute values scaled as in the total cross sections.  
Once again, we can verify that the s-wave dominates the production 
reaction up to $\sqrt{s} \lsim$ 2750. Recently, in
Ref.~\cite{Hanhart:2003xp}, the authors discussed the 
experimental methods and observable to determine the parity of the
$\Theta^+$ baryon 
with the polarized proton beam and target. They discussed the spin
correlation parameter $A_{xx}$ as well as cross sections. It is computed by
\be
A_{xx}=\frac{(^{3}\sigma{_0}+^{3}\sigma_{1})}{2\sigma_{0}}-1,
\ee
where $\sigma_{0}$ is the unpolarized total cross sections and the
polarized cross section are denoted as $^{2S+1}\sigma_{S_z}$. In
Fig.~\ref{axx} we present $A_{xx}$ including
$K^{*}$ exchange with and without the form factor.  As shown in the
figures $A_{xx}$ reflects very 
clearly the differences of the parity of $\Theta^+$. When the form
factor is included, the five cases of different $K^*$ 
coupling constants are similar and the resulting $A_{xx}$ fall into
well the region as indicated in Ref.~\cite{Hanhart:2003xp}. If the
from factor is not included,  there is an accidental
cancellation in the allowed s-wave amplitude for the (+,+) case and
the hence the p-wave 
contribution becomes significant at relatively low energy, which
changes the sign of $A_{xx}$ at $E_{\rm CM}\sim 2.75{\rm
GeV}$ for the positive parity case. However, very near the
threshold region, the sign of $A_{xx}$ 
is one as expected in the selection rule. In actual experiment, it is
necessary to detect $\Sigma$ also at the  
threshold region. It is worth mentioning that the quantity $A_{xx}$
does not depend very much on the less known parameters such as
coupling constants and form factors since their effects will be largely
cancelled when taking the rario of the two cross sections as shown in
Eq.~(\ref{axxeq}).  This advantage will give a chance to determine the 
parity of $\Theta^+$ without much theoretical ambiguities. Recently,
COSY-TOF collaboration announced that the experiment with 
polarized $pp$ scattering will be held in 2005~\cite{COSY-TOF}. We will look forward to see
a evidence to determine the parity of $\Theta^+$.

\section{Summary}
We have investigated the $\Theta^+$ production reactions via $\gamma
N$ and $NN$ scattering in the Born approximation. For the $\gamma N$ 
reaction, we considered two
different coupling schemes, pseudo-scalar (PS) and pseudo-vector (PV) couplings
for $KN\Theta^+$ vertex. We observed that the two coupling
schemes are not much different when the gauge invariant form factor is
employed since this form factor scheme enhances the $K$ and $K^*$
exchange contribution in the $t$--channel more than the $s$-- and the $u$--channels
which contain the whole difference between the PS and the PV
schemes. The reactions for the positive parity $\Theta^+$
showed about ten times larger total cross sections than those with the
negative one. This tendency is rather general for all reactions
including the following $NN$ scattering. However, it was 
difficult to determine the parity of $\Theta^+$ from $\gamma N$
reactions, by looking at, for instance, angular
distributions. As for the $NN$ scattering, we performed calculations
using the two different sets of interactions, the Nijmegen and
J\"ulich-Bonn potentials. The magnitudes of the total cross sections
of the two calculations were not so different 
from each other. However, the reactions with $\Sigma$ hyperon of the
J\"ulich-Bonn potential presented larger total cross sections than
those of the Nijmegen ones because of the larger cutoff mass of
J\"ulich-Bonn potential. Considering the measurement of
COSY-TOF~\cite{Abdel-Bary:2004ts}, the positive parity of $\Theta^+$ seems more
possible than the negative one qualitatively. However, the simple
comparison of the order of the total cross section does not include
sufficient information to determine the parity of $\Theta^+$
quantitatively. Therefore, we considered the model independent
way to determine the parity through the
polarized $pp$ scattering. We found clearly different behaviors of the
total cross sections for determination of the parity of $\Theta^+$
around the threshold region. The spin observable $A_{xx}$ which has
less theoretical ambiguities showed positive values for  
the negative parity of $\Theta^+$ and negative values for the positive parity
one. We have confirmed that the polarized $pp$ scattering will be a promising
method to determine the parity of $\Theta$. 

\section*{Acknowledgments}
We thank Tony Thomas, Ken Hicks, Hiroshi Toki, Kichiji Hatanaka,
Takashi Nakano, Tetsuo Hyodo and J.~K.~Ahn for fruituful discussions
and comments.  
The work of HCK is supported by the Korean Research Foundation
(KRF--2003--070--C00015). The work of SINam has been supported from
the scholarship endowed from the Ministry of Education, Science, 
Sports, and Culture of Japan.


\end{document}